\documentclass[a4paper,twocolumn,11pt,unpublished]{quantumarticle}

\pdfoutput=1
\usepackage[utf8]{inputenc}
\usepackage[english]{babel}
\usepackage[T1]{fontenc}
\usepackage{amsmath}
\usepackage{hyperref}

\usepackage{tikz}
\usepackage{lipsum}

\usepackage{braket}
\usepackage{amsfonts}
\usepackage{subcaption}
\usepackage{graphicx}
\usepackage{caption}

\usepackage{tikz}
\usetikzlibrary{quantikz}
\tikzset{operator/.append style={text width={width("$f_{t+1}$")}}}

\usetikzlibrary{shapes.geometric, arrows}
\tikzstyle{startstop} = [rectangle, rounded corners, minimum width=3cm, minimum height=1cm,text centered, draw=black, fill=red!30]
\tikzstyle{process} = [rectangle, rounded corners, minimum width=8cm, minimum height=0.7cm, text centered, draw=black, font = \small ]
\tikzstyle{process_s} = [rectangle, rounded corners, minimum width=2.2cm, minimum height=1cm, text centered, text width=1.5cm, draw=black ]

\usepackage{siunitx}
\usepackage[numbers,sort&compress]{natbib}

\usepackage{physics}

\DeclareUnicodeCharacter{2212}{-}
\usepackage{multicol}

\begin{document}

\title{Quantum Linear Multistep Method for Using a Quantum Oracle with Differential Equations}

\author{Kyoung Keun Park}
\affiliation{Department of Computer Science and Engineering, Seoul National University, 08826 Seoul, Republic of Korea}
\affiliation{Automation and System Research Institute, Seoul National University, 08826 Seoul, Republic of Korea}
\affiliation{NextQuantum, Seoul National University, 08826 Seoul, Republic of Korea}
\orcid{0009-0003-8112-7360}

\author{Kwangyeul Choi}
\affiliation{Department of Computer Science and Engineering, Seoul National University, 08826 Seoul, Republic of Korea}
\affiliation{Automation and System Research Institute, Seoul National University, 08826 Seoul, Republic of Korea}
\affiliation{NextQuantum, Seoul National University, 08826 Seoul, Republic of Korea}
\orcid{0009-0004-5308-8605}

\author{Minwoo Kim}
\affiliation{NextQuantum, Seoul National University, 08826 Seoul, Republic of Korea}
\affiliation{Department of Physics and Astronomy, Seoul National University, 08826 Seoul, Republic of Korea}
\orcid{0009-0007-6873-5299}

\author{Giwon Song}
\affiliation{NextQuantum, Seoul National University, 08826 Seoul, Republic of Korea}
\affiliation{Department of Electrical and Computer Engineering, Seoul National University, 08826 Seoul, Republic of Korea}
\orcid{0009-0007-7989-7891}

\author{Taehyun Kim}
\affiliation{Department of Computer Science and Engineering, Seoul National University, 08826 Seoul, Republic of Korea}
\affiliation{Automation and System Research Institute, Seoul National University, 08826 Seoul, Republic of Korea}
\affiliation{NextQuantum, Seoul National University, 08826 Seoul, Republic of Korea}
\affiliation{Institute of Computer Technology, Seoul National University, 08826 Seoul, Republic of Korea}
\affiliation{Inter university Semiconductor Research Center, Seoul National University, 08826 Seoul, Republic of Korea}
\affiliation{Institute of Applied Physics, Seoul National University, 08826 Seoul, Republic of Korea}
\orcid{0000-0003-3532-864X}
\email{taehyun@snu.ac.kr}
\maketitle


\begin{abstract}

Differential equations are a crucial mathematical tool used in a wide range of applications. If the solution to an initial value problem (IVP) can be transformed into an oracle, it can be utilized in various fields such as search and optimization, achieving quadratic speedup with respect to the number of candidates compared to its classical counterpart. In the past, attempts have been made to implement such an oracle using the Euler method. In this study, we propose a quantum linear multistep method (QLMM) that applies the linear multistep method, commonly used to numerically solve IVPs on classical computers, to generate a numerical solution of the IVP for use in a quantum oracle. We also propose a method to find the optimal form of QLMM for a given IVP. Finally, through computer simulations, we derive the QLMM formulation for an example IVP and show that the solution from the optimized QLMM can be used in an optimization problem.

\end{abstract}


\section{Introduction}

Numerous systems are described by ordinary differential equations (ODEs), and solving these equations is essential to predict their behavior over time \cite{0_0,0_1}. An initial value problem (IVP) involves finding a time-dependent solution starting from an initial value and progressing through the derivative of the variable, which is evaluated at each time step. While some differential equations allow for analytical solutions, the complexity of many real-world systems requires numerical approaches. Various numerical methods exist to solve IVPs, and selecting appropriate constants for these methods enables efficient solutions.

Quantum computers are expected to provide advantages over classical computers in tasks such as factorization \cite{0_2}, solving linear equations \cite{0_3,1_2}, and search problems \cite{2}. If a problem can be represented as a quantum oracle, quantum algorithms can achieve quadratic speed-ups in both search and optimization tasks over classical approaches \cite{3}. Developing a quantum oracle capable of solving IVPs could enable quantum-enhanced applications across diverse fields. To address differential equations with quantum computing, several quantum algorithms have been explored, such as the linear multistep method \cite{4_1} and Taylor series expansion \cite{1}, which offer exponential speed-up over classical methods, particularly as the system size of the differential equation grows. In these approaches, the solution is encoded in the amplitude of the quantum state, which poses challenges for using them within an oracle. Alternatively, an initial quantum approach to solving IVPs using the Euler method within a quantum oracle was proposed in Ref. \cite{4}, using fixed-point representation of variables.

When using the Euler method for solving differential equations, the truncation error is proportional to the square of the time step, making it relatively large compared to other numerical methods. By employing more advanced numerical methods, one can achieve lower errors with respect to the time step size, enabling fewer time steps to reach the same error level and thereby enhancing computational efficiency with respect to resources, such as qubits and circuit depth.

In this paper, we propose a quantum linear multistep method (QLMM) to efficiently solve IVPs for differential equations within a quantum oracle. We introduce the concept of QLMM and demonstrate its implementation in a quantum circuit using practical quantum gates. Our approach reduces the numerical error while minimizing the number of qubits at a given circuit depth, compared to the previous research \cite{4}. Additionally, we present an optimization technique for tailoring QLMM to different applications by framing the problem as a mixed-integer nonlinear programming (MINLP) problem.

The paper is structured as follows: Section 2 provides background on quantum oracles, IVP solutions, and quantum arithmetic. Section 3 describes the proposed QLMM algorithm, including its formulation and way to optimize the formulation of QLMM. In section 4, we demonstrate that an optimized QLMM implemented in a quantum oracle can effectively solve the target problem.

\section{Background Information}
Theoretical background relevant to the following discussion is provided in this section.

\subsection{Application of Quantum Oracle}
\paragraph{Quantum Search Algorithm using an Oracle}
Grover’s search algorithm efficiently identifies a solution that satisfies a quantum oracle, implemented on a quantum circuit, among a total of $C$ candidates in \begin{math} O(\sqrt{C} )\end{math} time \cite{2}. This problem, which requires $O(C)$ time on classical computers when candidates are unstructured, benefits from a quadratic speed-up on quantum computers. Grover's algorithm iteratively passes a superposition of candidate states through a quantum oracle and a diffuser. The oracle flips the sign of solutions within the superposition of candidates, while the diffuser amplifies the probability of states that correspond to potential solutions. By measuring the quantum state, a correct candidate is obtained with \begin{math} O(\sqrt{C} )\end{math} oracle calls. Enhanced algorithms based on Grover’s approach have been developed \cite{2_0, 2_0_1} and applied across various fields \cite{2_1, 2_2, 2_3}.

\paragraph{Quantum Optimization Algorithm using an Oracle}
Quantum oracles are also applicable to optimization problems, where the objective is to find a solution that minimizes a specific objective function among a set of $C$ feasible solutions. While a classical approach would require $O(C)$ time in the worst case, D\"{u}rr and H{\o}yer proposed an algorithm that uses a quantum oracle to find the solution in \begin{math} O(\sqrt{C}) \end{math} calls \cite{3}. Their algorithm iteratively sets a new threshold for the oracle and applies Grover's algorithm to identify candidates with an objective value below the threshold, eventually reaching the optimal solution. D\"{u}rr and H{\o}yer's algorithm has been applied in various applications related to optimization \cite{3_1, 3_2}.


\subsection{Problem Inside the Oracle}
\subsubsection {Differential Equations}

\paragraph {Ordinary Differential Equation}
An ordinary differential equation involves derivatives with respect to a single variable, and may include higher-order derivatives. Systems of ODEs can arise when an equation involves multiple variables, or when a higher-order equation is converted into a system of first-order equations \cite{5_2}.

\paragraph {Initial Value Problem}
An initial value problem requires finding a time-dependent solution that satisfies an initial condition and the governing differential equation. Mathematically, an IVP can be expressed as
\begin{equation}\label{IVP}
    \dot{\vb*{x}}(t) = \vb*{f}(t,\vb*{x}(t)), t\in[t_0,t_f], \vb*{x}(t_0) =\vb*{x}_0,
\end{equation}

\noindent where \begin{math} \vb*{x}(t)\in \mathbb{R}^D\end{math} which implies that the ODEs are dealing with $D$ variables.

For complex cases where an analytical solution is unattainable, a numerical method with a fixed time step $h$ is used to compute $\vb*{x}_n$, the estimated value at \begin{math} t_n=t_0 + nh \end{math}, iteratively based on \begin{math} \vb*{f}_n = \vb*{f}(t_n, \vb*{x}_n)\end{math} until $t_f$ is reached, corresponding to step $N = \lfloor \frac{t_f - t_0}{h}\rfloor $.

\paragraph{Euler Method}
The Euler method is a straightforward approach for solving IVPs, with the value at the $(n+1)$th time step given by
\begin{equation}\label{Euler}
    \vb*{x}_{n+1} = \vb*{x}_n + h \vb*{f}_n.
\end{equation}

The local truncation error of the Euler method is \begin{math} O(h^2) \end{math}. To achieve lower truncation errors, methods such as the linear multistep method or Runge-Kutta method are employed. The Euler method can be considered a special case of both methods \cite{5_0}.

\subsubsection {Linear Multistep Method}
The linear multistep method (LMM) computes the next time step by using a weighted sum of previous values, leveraging already computed results. An explicit $k$-step LMM has the general form
\begin{equation}\label{LMM_scalar}
    x_{n+k}+\sum_{i=0}^{k-1}\alpha_i x_{n+i} = h \sum_{j=0}^{k-1}\beta_j f_{n+j},
\end{equation}

\noindent where $\alpha_i$ and $\beta_j$ are constants chosen to satisfy specific convergence conditions \cite{5_0}.

A method is convergent if the numerical solution approaches the actual solution as the time step size $h$ approaches zero. For an LMM, convergence requires consistency and zero-stability \cite{5_0}. An operator \begin{math} \mathcal{R}_h \end{math} is defined by
\begin{multline}\label{LDO}
    \mathcal{R}_h x(t) = x(t+kh) \\
    + \sum_{j=0}^{k-1} {(\alpha_j x(t+jh) - h \beta_j x'(t+jh))},
\end{multline}

\noindent which is related to the truncation error. If this expression is \begin{math}O(h^{p+1})\end{math} for any continuous, differentiable function $x(t)$, the method is consistent of order $p$, and for $p>0$, the LMM is considered consistent \cite{5_0}. Moreover, an LMM is zero-stable if all roots of
\begin{equation}\label{rho}
\rho(r) = r^k + \alpha_{k-1}r^{k-1} + \cdots +\alpha_0=0,
\end{equation}
\noindent have absolute values less than 1, except for one root which can have an absolute value of 1. If the combination of \begin{math} \alpha_i \end{math} and  \begin{math}\beta_j\end{math} satisfies certain conditions, the LMM can be convergent. According to the first Dahlquist barrier,  \begin{math} p \leq k \end{math} to ensure that the error with respect to the time step size is bounded \cite{5_1_1}.

The time step $h$ must remain reasonably small to avoid excessive iterations. For the IVP given by 
\begin{equation}\label{diag}
\dot{x}(t)=\lambda x(t), x(0) = x_0,
\end{equation}

\noindent where $\lambda\in \mathbb{C}$ with a negative real part, and $x_0$ is an arbitrary value, an LMM is absolutely stable if the solution converges to 0 as time approaches infinity. An LMM is absolutely stable  when all roots of 
\begin{equation}\label{sigma}
    \rho(r) -h \lambda (r^k + \beta_{k-1}r^{k-1} + \cdots +\beta_0) = 0
\end{equation}

\noindent should have absolute values less than 1, except for one root which can have an absolute value of 1. Absolute stability of a 2-step LMM is achieved if the Jury condition (Eq.~\ref{2stable}) is satisfied.
\begin{align}
\alpha_0 &<1 \nonumber\\
1+(\alpha_1-\lambda h\beta_1)+\alpha_0-\lambda h\beta_0 &>0 \label{2stable}\\
1-(\alpha_1-\lambda h\beta_1)+\alpha_0-\lambda h\beta_0 &>0 \nonumber
\end{align}

\noindent Depending on the combination of \begin{math} \alpha_i \end{math} and \begin{math}\beta_j\end{math}, the time step \begin{math} h \end{math} may be chosen freely for absolute stability. However, according to the Dahlquist second barrier theorem, no explicit LMM can have absolute stability with arbitrary time step \begin{math} h \end{math} \cite{5_1_2}.

An LMM incurs truncation error due to non-zero time steps. The local truncation error can be analyzed using a Taylor expansion of each term in the LMM. In a $k$-step LMM, initial values up to time step $(k-1)$ are required, and both initial error and truncation error influence accuracy \cite{5_3}.

Additionally, for a vector variable $\vb*{x}\in \mathbb{R}^D$, the LMM can also be applied to systems of ODEs, such as 
\begin{equation}\label{LMM}
    \vb*{x}_{n+k}+\sum_{i=0}^{k-1}\alpha_i \vb*{x}_{n+i} = h \sum_{j=0}^{k-1}\beta_j \vb*{f}_{n+j}.
\end{equation}

\subsubsection {Runge-Kutta Method}
The Runge-Kutta (RK) method computes \begin{math} \vb*{x}_{n+1} \end{math} from \begin{math} \vb*{x}_n \end{math} using intermediate values, with the $p$-order RK method achieving a local truncation error of \begin{math} O(h^{p+1}) \end{math} \cite{5_0}. The commonly used 4th-order RK method for computing \begin{math} \vb*{x}_{n+1} \end{math} is given by
\begin{align}\label{RK4}
\vb*{x}_{n+1} &= \vb*{x}_n + \frac{h}{6} (\vb*{k}_{n,1} + 2\vb*{k}_{n,2}+2\vb*{k}_{n,3}+\vb*{k}_{n,4}) \nonumber\\
\vb*{k}_{n,1}&=\vb*{f}(t_n,\vb*{x}_n)  \\
\vb*{k}_{n,2}&=\vb*{f}(t_n+\frac{h}{2},\vb*{x}_n+\frac{h}{2} \vb*{k}_{n,1}) \nonumber\\
\vb*{k}_{n,3}&=\vb*{f}(t_n+\frac{h}{2},\vb*{x}_n+\frac{h}{2} \vb*{k}_{n,2}) \nonumber\\
\vb*{k}_{n,4}&=\vb*{f}(t_n+h,\vb*{x}_n+h\vb*{k}_{n,3}) \nonumber
\end{align}

\noindent where $\vb*{k}_{n,1}, \vb*{k}_{n,2}, \vb*{k}_{n,3}, \vb*{k}_{n,4}$ are intermediate derivatives.


\subsection {Quantum Arithmetic}
Quantum arithmetic involves performing calculations on quantum circuits. To apply LMM in quantum computing, quantum arithmetic dealing with integers and real numbers must be considered.

\subsubsection {Quantum Integer Arithmetic}
For integers \begin{math} n_a \end{math} and \begin{math} n_b \end{math} that can be represented by quantum registers composed of $N_z$ qubits, the operation \begin{math} \ket{n_a}\ket{n_b} \rightarrow \ket{n_a}\ket{n_a + n_b \mod{2^{N_z}}} \end{math} can be performed using the quantum Fourier transform \cite{7_3}. In this paper, quantum integer arithmetic for summation and multiplication proposed in Ref. \cite{7_3} is used to calculate LMM.

\subsubsection {Quantum Real Number Arithmetic}\label{2_3_2}
In general calculations, real numbers also need to be processed in addition to integers. A floating point number can be represented by decomposing it into a sign, mantissa, and exponent in binary format, similar to classical computer calculations. For quantum states representing two real numbers \begin{math} r_a\end{math} and \begin{math} r_b\end{math}, the sum (e.g., \begin{math} \ket{r_a}\ket{r_b}\ket{0} \rightarrow \ket{r_a}\ket{r_b}\ket{r_a+r_b} \end{math}) or product (e.g., \begin{math} \ket{r_a}\ket{r_b}\ket{0} \rightarrow \ket{r_a}\ket{r_b}\ket{r_a\times r_b} \end{math}) can be computed using additional qubits \cite{7_0}. Since addition and multiplication of floating-point numbers are possible in qubits other than those storing the original values \begin{math} r_a\end{math} and \begin{math} r_b\end{math}, subtraction and division \cite{7_2}, as well as other operations, can be computed as long as additional qubits are available. Further operations, such as multiplication by powers of 2 (e.g., \begin{math} \ket{r_a}\ket{n_b}\rightarrow\ket{r_a\times 2^{n_b}}\ket{n_b} \end{math}), are also possible by modifying the exponent alone.

Note that the algorithm proposed in this research employs two different types of arithmetic: the method summarized in this section \cite{7_0} and the approach detailed in Appendix~\ref{ap1}. More details are discussed in section~\ref{3_1_3}.

\section{Algorithm}

In this section, we propose an algorithm for efficiently solving IVP within a quantum oracle.

\subsection {Quantum Linear Multistep Method}

\subsubsection{Idea of QLMM}
Euler's method, linear multistep method, and Runge-Kutta method are well-known classical algorithms for numerically solving initial value problems. Since every classical gate can be implemented using a quantum gate, every classical calculation can also be performed on a quantum computer, provided there are sufficient qubits. Solving an IVP using numerical methods involves calculating a new solution for the next time step and storing it in additional memory. To use qubits efficiently in solving IVPs, uncomputation must be applied to reset intermediate qubits. However, since quantum circuits must be reversible, qubit information cannot be directly erased. We propose the quantum linear multistep method (QLMM), which implements the LMM on quantum circuits. Unlike the Runge-Kutta method, QLMM uses previous time-step information to iteratively update values. The general form of QLMM is
\begin{equation}\label{QLMM}
    \vb*{y}_{n+k}+\sum_{i=0}^{k-1}\alpha_i \vb*{y}_{n+i} = h \sum_{j=0}^{k-1}\beta_j \vb*{f}_{n+j},
\end{equation}

\noindent for $n=0$ to $N$. QLMM requires \begin{math} \alpha_i\end{math} and \begin{math} \beta_j \end{math} that meet the consistency and zero-stability conditions of classical LMM. A $k$-step QLMM evolves values from the $n$th to the $(n+k-1)$th time step into values from the $(n+1)$th to the $(n+k)$th time step using a unitary operator $U_n$, as
\begin{align}
U_n\bigotimes_{i=0}^{k-1}\ket{\vb*{y}_{n+i}} \ket{0}^{\otimes R_c}\ket{0}^{\otimes R_u}\nonumber \\
 =\bigotimes_{j=1}^{k}\ket{\vb*{y}_{n+j}}  \ket{etc}\ket{0}^{\otimes R_u} \label{U_eq}
\end{align}
where $R_c$ is the number of auxiliary qubits that cannot be uncomputed, and $R_u$ is the number of auxiliary qubits that can be uncomputed.

Suppose an arbitrary function $u(a)$, with variable $a$, has been implemented on a quantum circuit to perform the quantum operation \begin{math}\ket{a}\ket{0} \rightarrow \ket{a}\ket{u(a)}\end{math}. The value $\ket{a}\ket{0}$ can be recovered from $\ket{a}\ket{u(a)}$ through uncomputation by using $\ket{a}$. In the case of the Euler method, if a state \begin{math} \ket{\vb*{y}_n}\ket{0}\end{math} at a specific time step $n$ is transformed into \begin{math} \ket{\vb*{y}_{n+1}}\ket{u(\vb*{y}_{n+1})} \end{math} using a unitary operator, uncomputing the state $\ket{u(\vb*{y}_{n+1})} $ becomes challenging because \begin{math} \ket{\vb*{y}_n}\end{math} has already been transformed into \begin{math} \ket{\vb*{y}_{n+1}}\end{math}. However, QLMM enables efficient differential equation solving on quantum circuits by performing uncomputation using information stored in $ \ket{\vb*{y}_{n+1}}$ through $\ket{\vb*{y}_{n+k-1}} $ when updating \begin{math} \ket{\vb*{y}_{n}} \end{math} to \begin{math} \ket{\vb*{y}_{n+k}} \end{math}.

\subsubsection{Basic Structure of QLMM}\label{3_1_3}

\paragraph{Floating-Point Number Expression in QLMM}

Dealing with $\vb*{x_n}$ having $D$ different elements, each element can be represented by a different number of qubits because the required precision can vary. QLMM stores information in floating-point format, expressing information using quantum registers for the mantissa and exponent. Each quantum register consists of a different number of qubits: \begin{math} \vb*{M} \in \mathbb{N}^D  \end{math} qubits for the mantissa and \begin{math} \vb*{E} \in \mathbb{N}^D  \end{math} qubits for the exponent. To avoid negative values, each  \begin{math} \vb*{x}_n \end{math} is adjusted with a constant $\vb*{v}$ so that $\vb*{y}_n=\vb*{x}_n+\vb*{v}>0$ for $n=0,\cdots,N$. The consistency of $k$-step QLMM remains satisfied when converting \begin{math} \vb*{x}_n \end{math} to \begin{math} \vb*{y}_n \end{math}. Also the $d$th element of $\vb*{y}_n$ will be denoted as $y_{n,d}$.

\paragraph{Calculation Method of QLMM}
QLMM can be implemented in a quantum circuit by the following procedures: First, multiply $\vb*{y}_{n}$ stored in a quantum register by $-\alpha_0$ to obtain $-\alpha_0 \vb*{y}_{n}$. For a floating-point number $\ket{y_{n,d}}$, the operation $\ket{y_{n,d}}\rightarrow\ket{2^{a_0}\times y_{n,d}}$, where \begin{math} a_0 \in \mathbb{Z} \end{math}, can be easily performed by adding $a_0$ to the qubits representing the exponent of $y_{n,d}$. Because the solution of Eq.~\eqref{sigma} must have an absolute value smaller than 1, the absolute value of $\alpha_0$ must also be smaller than 1, assuming $\beta_0 = 0$. Additionally, since the result of the calculation $-\alpha_0 \vb*{y}_n$ is desired to be positive, it follows that $-\alpha_0 > 0$. Therefore, \begin{math} \alpha_0 \end{math} in our QLMM must satisfy: 
\begin{equation}\label{a0}
\alpha_0=-2^{-a_0},a_0\in \mathbb{N}.
\end{equation}

Next, compute the value of \begin{math} -\sum_{i=1}^{k-1}\alpha_i \vb*{y}_{n+i} + h \sum_{j=1}^{k-1}\beta_j \vb*{f}_{n+j}\end{math}. To accomplish this, retrieve the bias-adjusted values $\vb*{x}_{n+1}$ through $\vb*{x}_{n+k-1}$ from $\vb*{y}_{n+1}$ through $\vb*{y}_{n+k-1}$. Then, calculate values $ \vb*{f}_{n+1} $ through $ \vb*{f}_{n+k-1} $ using the floating-point summation and multiplication method summarized in section~\ref{2_3_2} with auxiliary qubits \cite{7_0}. By a weighted sum of these values, \begin{math} -\sum_{i=1}^{k-1}\alpha_i \vb*{y}_{n+i} + h \sum_{j=1}^{k-1}\beta_j \vb*{f}_{n+j}\end{math} can be computed.

The value calculated from the previous step is added to \begin{math} -\alpha_0 \vb*{y}_n \end{math} to calculate \begin{math} \vb*{y}_{n+k}\end{math}. When adding two arbitrary floating-point numbers \begin{math}r_a\end{math} and \begin{math} r_b\end{math}, existing quantum floating-point arithmetic can be used if enough additional auxiliary qubits are available. If a summation of floating-point numbers is desired to be performed without the help of any additional qubits, such as $\ket{r_a}\ket{r_b}\rightarrow\ket{r_a}\ket{r_a+r_b} $, this addition may not be performed when \begin{math}r_a>r_b\end{math} since all operations in quantum circuits must be reversible. Even with summation between similarly sized numbers, overflow can occur during summation. Therefore, in general, summation should be performed using additional qubits, like \begin{math} \ket{r_a}\ket{r_b}\ket{0}^{\otimes S_c}\ket{0}^{\otimes S_u} \rightarrow \ket{r_a+r_b}\ket{r_b}\ket{etc}\ket{0}^{\otimes S_u} \end{math}. The difference between the additional qubits is that the $S_u$ qubits are uncomputed back to their original value of 0, while the $S_c$ qubits hold arbitrary values. The smaller the $S_c$, the fewer total qubits are required for the entire numerical computations, as $S_c$ qubits are consumed at each time step. One possible summation implementation is explained in Appendix~\ref{ap1}.

\begin{figure*}[ht]
\begin{center}
\begin{quantikz}[row sep=0.2cm]
\lstick{$\ket{0}$} & \qw&\qw&\qw&\qw& \gate[2][0.3cm]{g_3}& \qw & \qw & \rstick{$\ket{etc}$}\qw\\
\lstick{$\ket{\vb*{y}_n}$} & \gate{g_1} & \qw &\qw&\qw&\qw& \qw & \gate[4]{g_5} & \rstick{$\ket{\vb*{y}_{n+1}}$}\qw\\
\lstick{$\ket{\vb*{y}_{n+1}}$} & \ctrl{3} &\ \ldots\ \qw & \qw & \ctrl{2} &\qw& \gate[7]{g_4} &  & \rstick{$\ket{\vb*{y}_{n+2}}$}\qw\\
\wave&&&&&&&&&&&&\\
\lstick{$\ket{\vb*{y}_{n+k-1}}$} &\qw & \ \ldots\ \qw&  \ctrl{3} & \ctrl{1} &\qw & &  & \rstick{$\ket{\vb*{y}_{n+k}}$}\qw\\
\lstick{$\ket{0}$} & \gate[style={inner xsep=5pt}]{\vb*{f}_{n+1}} & \ \ldots\ \qw &\qw & \ctrl{2} &\qw&  & \qw & \rstick{$\ket{0}$}\qw\\
\wave&&&&&&&&&&&&\\
\lstick{$\ket{0}$} &\qw & \ \ldots\ \qw & \gate[style={inner xsep=8pt}]{\vb*{f}_{n+k-1}} & \ctrl{1} &\qw&  & \qw & \rstick{$\ket{0}$}\qw\\
\lstick{$\ket{0}$} &\qw&\qw&\qw& \gate{g_2} & \ctrl{-8} &  & \qw & \rstick{$\ket{0}$}\qw
\end{quantikz}
\end{center}
\caption{Quantum circuit implementation of $U_n$: one time step calculation for QLMM. $ g_1$ : set of gates that multiply $-\alpha_0$ by $\vb*{y_n}$, $g_2$ : set of gates that perform weighted sum to calculate $-\sum_{i=1}^{k-1}\alpha_i \vb*{y}_{n+i} + h \sum_{j=1}^{k-1}\beta_j \vb*{f}_{n+j}$, $g_3$ : set of gates that perform summation between $-\alpha_0\vb*{y_n}$ and $-\sum_{i=1}^{k-1}\alpha_i \vb*{y}_{n+i} + h \sum_{j=1}^{k-1}\beta_j \vb*{f}_{n+j}$, $g_4$ : set of gates that uncompute qubits to their initial state, $g_5$ : set of gates that rearrange states in order from $n+1$ to $n_k$}
\label{QLMM_fig}
\end{figure*}
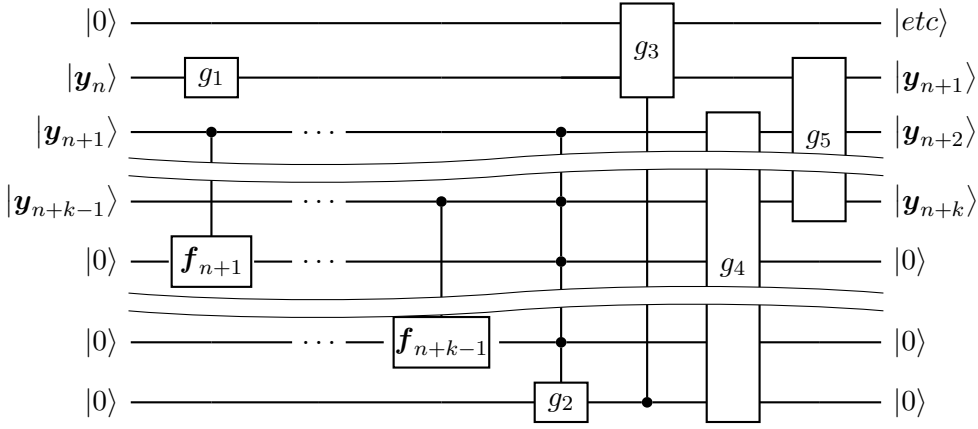

Next, the uncomputation of $ \vb*{f}_{n+i} $ for $i=1,\cdots,k-1$ is performed using $ \vb*{y}_{n+i}$. Since $ \vb*{f}_{n} $ cannot be uncomputed due to the absence of remaining information about $ \vb*{y}_{n} $, $\beta_0$ must be set to 0 to ensure that $ \vb*{f}_{n} $ is not used in the calculation. A swap operation then rearranges the states in the order \begin{math} \ket{\vb*{y}_{n+1}}\ket{\vb*{y}_{n+2}}\cdots \ket{\vb*{y}_{n+k}} \end{math}. The entire process is illustrated in Fig.~\ref{QLMM_fig}.

\subsubsection {Use of Quantum LMM}

\begin{figure*}[ht] 
    \begin{subfigure}{0.5\textwidth}
        \begin{center}
        \begin{quantikz}[slice style=blue]
        \lstick{$\ket{c}_{Q_c} $} &\ctrl{1}        &\ctrl{1}       &\ctrl{1}    &\qw           &\ctrl{1}  &\rstick{$\ket{c}_{Q_c} $}\qw\\
        \lstick{$\ket{0}_{Q_y} $} &\gate[2]{x_0^{[c]}}&\gate[2]{g_1}&\gate[2]{g_{2a}}\slice{Eq.~\eqref{tf_oracle}} &\gate[3]{g_3}&\gate[2]{g_4}      &\rstick{$\ket{0}_{Q_y}$}\qw\\
        \lstick{$\ket{0}_{Q_a}$} &                &               &            &              &               &\rstick{$\ket{0}_{Q_a}$}\qw\\
        \lstick{$\ket{0}_{Q_o}$} &\qw             &\qw            &\qw         &              &\qw            &\rstick{$\ket{g_5}_{Q_o}$}\qw
        \end{quantikz}
        \end{center}
        \caption{}
        \label{final_time}
    \end{subfigure}%
    \begin{subfigure}{0.5\textwidth}
        \begin{center}
        \begin{quantikz}[]
        \lstick{$\ket{c}_{Q_c}$} &\ctrl{1}        &\ctrl{1}       &\ctrl{1}    &\qw           &\ctrl{1}  &\rstick{$\ket{c}_{Q_c}$}\qw\\
        \lstick{$\ket{0}_{Q_y}$} &\gate[2]{x_0^{[c]}}&\gate[2]{g_1}&\gate[2]{g_{2b}}\slice{Eq.~\eqref{t_oracle}} &\gate[3]{g_3}&\gate[2]{g_4}     &\rstick{$\ket{0}_{Q_y}$}\qw\\
        \lstick{$\ket{0}_{Q_a}$} &                &               &            &              &               &\rstick{$\ket{0}_{Q_a}$}\qw\\
        \lstick{$\ket{0}_{Q_o}$} &\qw             &\qw            &\qw         &              &\qw         &\rstick{$\ket{g_5}_{Q_o}$}\qw\\
        \lstick{$\ket{n}_{Q_n}$} &\qw             &\qw            &\ctrl{-2}   &\qw           &\ctrl{-2}          &\rstick{$\ket{n}_{Q_n}$}\qw
        \end{quantikz}
        \end{center}
        \caption{}
        \label{every_time}
    \end{subfigure}%
    \caption{(a) Oracle implementation using QLMM that uses the solution at the final time (b) Oracle implementation using QLMM that uses the solution at every time step. $ g_1$ : set of gates that calculate initial values using RK method, $g_{2a}$ : set of gates that apply the time-dependent QLMM sequentially up to time step $N$, $g_{2b}$ : set of gates that apply the time-independent QLMM $n$ times, $g_3$ : set of gates that express quantum oracle for a specific problem, $g_4$ : set of gates that uncompute previous calculations except for those performed by $g_3$, $g_5$ : quantum state representing the quantum oracle result. }
\end{figure*}
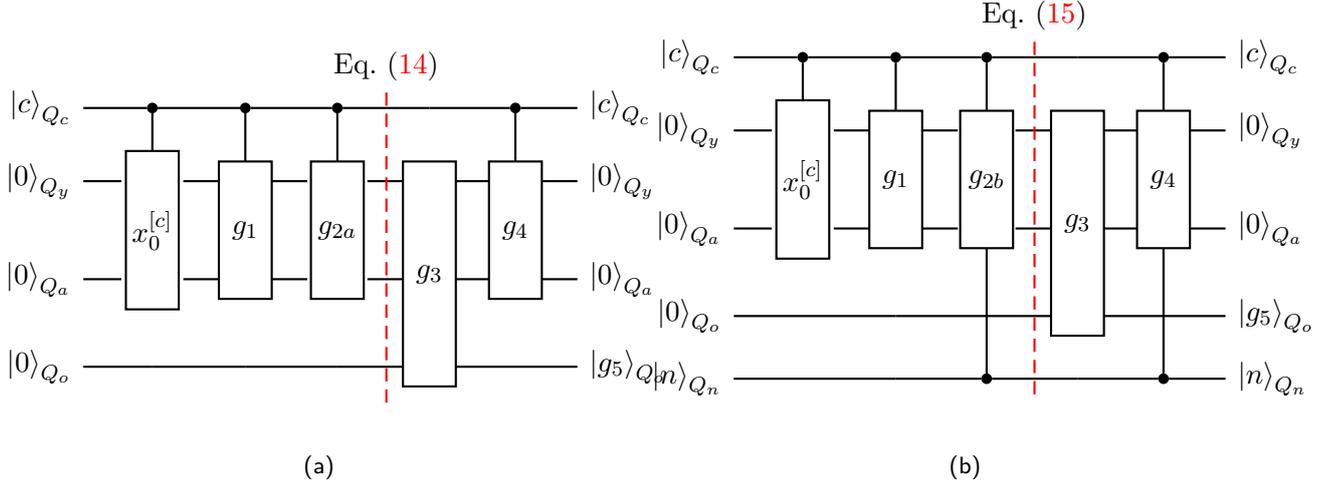
\paragraph {Using QLMM as part of a Quantum Algorithm}

Like LMM, QLMM requires initial values for certain time steps from $0$ to $(k-1)$ to begin calculations. If the initial value is prepared only for $\vb*{x}_{0}$, the subsequent values from time steps $1$ to $(k−1)$ must be calculated without QLMM. If this is done using the Euler method, the solutions will contain significant initial errors. Therefore, QLMM sets values from $\vb*{x}_{1}$ to $\vb*{x}_{k-1}$ using the Runge-Kutta method, which yields smaller errors compared to the Euler method.

When applying a $k$-step QLMM within a quantum algorithm, the process proceeds as follows. First, create a superposition of the parameters of the problem. Next, generate the initial value \begin{math} \ket{\vb*{y}_0}\end{math} according to the problem parameters. Then, generate the subsequent terms \begin{math} \ket{\vb*{y}_1}\end{math} through \begin{math} \ket{\vb*{y}_{k-1}}\end{math} on separate quantum register using the Runge-Kutta method. Then, perform the QLMM over time to perform calculations, and apply the oracle to the results. Finally, uncompute the intermediate results used in the computation. When applying QLMM to a situation with $C$ different superpositions, there are two different problems that QLMM can approach:

\paragraph {Using Solution at Final Time for Oracle}
In some cases, the solution of the IVP at a specific time $t_f$ is used within an oracle. These types of problem can be solved using QLMM, as shown in Fig.~\ref{final_time}. QLMM is computed up to the desired final time $t_f$, resulting in the quantum state
\begin{equation}\label{tf_oracle}
    \frac{1}{\sqrt C}\sum_{c=0}^{C-1} \ket {c}_{Q_c} \ket {\vb*{y}_{N}^{[c]},\cdots,\vb*{y}_{N+k-1}^{[c]}}_{Q_y} \ket{etc}_{Q_a},
\end{equation}
\noindent where the oracle is applied to the value at time step $N$, represented by \begin{math}\vb*{y}_{N}^{[c]} \end{math}. Using QLMM in this setup allows a quantum computer to solve search or optimization problems with a computational complexity of $ O(\sqrt{C}N)$ if an IVP with a time-dependent differential equation can be efficiently prepared.

\paragraph {Using Solution at all Time Steps for Oracle}
For certain oracle-based problems, the solution of the IVP at various time steps is required. These types of problems can be solved using QLMM, as shown in Fig.~\ref{every_time}. In these cases, a superposition of all combinations and all time steps from 0 to $N$ is created. QLMM is then applied at each time step, resulting in
\begin{align}\label{t_oracle}
\frac{1}{\sqrt {CN}} \sum_{c=0}^{C-1} &\ket {c}_{Q_c}  \sum_{n=0}^{N} \ket {n}_{Q_n} \\ \nonumber
&\ket {\vb*{y}_{n}^{[c]},\cdots,\vb*{y}_{n+k-1}^{[c]}}_{Q_y}  \ket{etc}_{Q_a}
\end{align}

\noindent The oracle is then applied to \begin{math}\vb*{y}_{n}^{[c]} \end{math}, the solution of the IVP at time step $n$ for combination \begin{math}c\end{math}. By allowing values at different times and combinations to occupy the same quantum register, QLMM enables efficient use of quantum algorithms, resulting in computational complexity of \begin{math} O(\sqrt{C}N^{3/2})\end{math} for search and optimization tasks. This is possible only for an IVP with a time-independent differential equation, and the problem must be expressed efficiently in a quantum circuit.


\subsection {Mathematical Formulation of QLMM}
A $k$-step QLMM can be applied, with feasible forms described as follows.

\subsubsection {1-step QLMM}
The general form of the explicit 1-step LMM corresponds to the Euler method, as shown in Eq.~\eqref{Euler}. This method calculates $\ket{\vb*{x}_{n+1}}$ and overwrites it in the same quantum register that contains $\ket{\vb*{x}_{n}}$. Since there is no information about $\vb*{x}_n$, the intermediate components, such as $\vb*{f}_n$, cannot be uncomputed, making the 1-step QLMM impractical.

\subsubsection {2-step QLMM}
The simplest form of QLMM is the 2-step QLMM, which iteratively computes values using
\begin{equation}\label{2QLMM}
    \vb*{y}_{n+2}+\alpha_1 \vb*{y}_{n+1}+\alpha_{0}\vb*{y}_{n}=h \beta_1 \vb*{f}_{n+1}.
\end{equation}

\noindent Although a 2-step LMM can achieve second-order consistency, the 2-step QLMM attains only first-order consistency due to the constraints of coefficient
\begin{align}\label{2consistent}
1+\alpha_1+\alpha_0&=0, \\
2+\alpha_1&=\beta_1. \nonumber
\end{align}

\noindent To satisfy the stability condition, the 2-step QLMM must meet the Jury condition and fulfill Eq.~\eqref{2stable} with $\beta_0=0$.

\subsubsection {3-step QLMM}
The 3-step QLMM improves accuracy over the 2-step QLMM, although it requires additional computational effort. Its general form is
\begin{multline}\label{3QLMM}
    \vb*{y}_{n+3} + \alpha_2 \vb*{y}_{n+2}+\alpha_1 \vb*{y}_{n+1}+\alpha_{0}\vb*{y}_{n}\\
    =h( \beta_2 \vb*{f}_{n+2}+\beta_1 \vb*{f}_{n+1}).
\end{multline}

\noindent While the 3-step LMM can achieve third-order consistency, the 3-step QLMM attains up to second-order consistency, requiring the conditions
\begin{align}
1+\alpha_2 + \alpha_1+\alpha_0&=0 \nonumber,\\
3+2\alpha_2 + \alpha_1&=\beta_2 + \beta_1 \label{3consistent},\\
9/2+2\alpha_2 + 1/2 \alpha_1&= 2\beta_2 + \beta_1. \nonumber
\end{align}

\noindent To meet the stability condition, the absolute values of solutions in
\begin{equation}\label{3stable}
r^3 + (\alpha_2 - h\beta_2) r^2 + (\alpha_1 - h \beta_1)r + \alpha_0= 0
\end{equation}

\noindent should have absolute values less than 1, except for one root which can have an absolute value of 1.

\subsubsection {Other QLMMs}
Using more past time steps (a larger $k$) in the calculation with QLMM can theoretically result in a smaller local truncation error through numerical calculations. The general form of a $k$-step QLMM is given by Eq.~\eqref{QLMM}. According to the first Dahlquist barrier, the error of a $k$-step QLMM is \begin{math}O(h^{p+1})\end{math}, and since \begin{math} p\leq k \end{math}, the error can be reduced by choosing appropriate \begin{math} \alpha_i \end{math} and \begin{math} \beta_j \end{math}. To satisfy consistency,
\begin{align}
1+\sum_{i=0}^{k-1} \alpha_i&=0, \nonumber\\
k^m + \sum_{i=1}^{k-1} i^m \alpha_i &= m\sum_{j=1}^{k-1} j^{m-1} \beta_j, \label{kconsistent}
\end{align}

\noindent for $m=1,\cdots,m_{max}$ must hold, where $m_{max}$ needs to be as large as possible to maintain a low truncation error. Solving a $k$-order polynomial, as shown in Eq.~\eqref{sigma}, meets the stability condition, though calculations become challenging for \begin{math} k \geq 5 \end{math}.


\subsection {Optimization of QLMM Parameters} \label{3_3}
QLMM can be designed by specifying the variables, time steps, and number of qubits required for each calculation. To efficiently execute numerical calculations in a quantum circuit, an appropriate time step and an optimal number of qubits for representing values should be carefully chosen. Depending on the optimization goal, various objective functions can be defined to optimize the form of QLMM. In the following discussion, the vectors \begin{math} \vb*{u}\in \mathbb{R}_{>0}^D\end{math} and \begin{math}\vb*{l}\in \mathbb{R}_{<0}^D\end{math} represents the upper and lower bounds of \begin{math}\vb*{f}(t,\vb*{x})\end{math} in every possible combination that the quantum algorithm is considering between time $t_0$ and $t_f$. In the following subsections, the problem formulation, objective function, and constraints of the optimization problem for finding the optimal form of QLMM will be discussed.

\subsubsection {Problem Formulation of Optimized QLMM}
\paragraph{Non-convex Mixed Integer Nonlinear Programming}

A mixed-integer nonlinear programming (MINLP) problem involves an objective function and constraints that are nonlinear, with some variables constrained to be integers. The problem is defined as
\begin{equation}\label{MINLP}\begin{split}
     \textrm{minimize } & o_0(\vb*{s}_Z, \vb*{s}_R)\\
     \textrm{subject to }& o_i(\vb*{s}_Z, \vb*{s}_R)\leq 0, i=1,\cdots,n_o
 \end{split}\end{equation}

\noindent where $\vb*{s}_R\in \mathbb{R}_{+}^{n_R}$ and $\vb*{s}_Z\in \mathbb{Z}_{+}^{n_Z}$, $n_o$ refers to the number of constraints, $n_R$ refers to the number of continuous variables, and $n_Z$ refers to the number of integer variables \cite{9_10}. For convex functions $o_i(\vb*{s}_Z, \vb*{s}_R)$, the problem qualifies as convex optimization, which is simpler to solve \cite{9_6}. However, non-convex MINLP problem \cite{9_7} are NP-hard, as they cannot easily be solved \cite{9}.

\paragraph{Variables of Optimized QLMM}
Optimization variables include mantissa qubit count \begin{math}\vb*{M}\end{math}, exponent qubit count \begin{math}\vb*{E}\end{math}, bias \begin{math}\vb*{v}\end{math}, QLMM coefficients \begin{math}\vb*{\alpha}\end{math} and \begin{math} \vb*{\beta} \end{math}, time step size \begin{math} h \end{math}, and extra variables related to quantum circuit implementation. Since the number of qubits must be a natural number, an integer constraint is applied.

\subsubsection{Objective Functions of Optimized QLMM}
\paragraph{Minimal Number of Qubits}
In practical quantum computing, using many qubits is challenging, and quantum computers with large numbers of qubits are likely a distant prospect. Using fewer qubits is more efficient and offers greater possibility for implementing a quantum algorithm to solve the IVP problem. To use as few qubits as possible while ensuring the algorithm's solution maintains an error below $\epsilon$, the objective function sums the number of qubits used to implement QLMM in the quantum circuit. The mathematical expression of this objective function can be described as
\begin{equation}\label{Low}
k\sum_{d=0}^{D-1}(M_d+E_d) + N R_c + R_u.
\end{equation}

\noindent If $R_c$ and $R_u$ can be modeled as linear functions with respect to the optimization variables, then the objective function becomes a quadratic function with integer variables.

\paragraph{Faster Calculation}
One of the primary goals of using quantum algorithms is to achieve faster problem-solving capabilities. Therefore, if the same problem can be solved more quickly, it is considered a better QLMM. The computation speed of a quantum algorithm using a quantum oracle is influenced by the problem size and the depth of the oracle. Since the QLMM must be applied repeatedly up to the final time step to implement the quantum oracle, the objective function must address both the number of time steps and the depth of QLMM while maintaining an error below $\epsilon$. The mathematical expression of this objective function can be described as
\begin{equation}\label{Fast}
N q_k(\vb*{M},\vb*{E},R_c),
\end{equation}

\noindent where \begin{math} q_k(\vb*{M},\vb*{E},R_c) \end{math} represents the depth of the quantum circuit in Fig.~\ref{QLMM_fig} for a $k$-step QLMM. If \begin{math} q_k(\vb*{M},\vb*{E},R_c) \end{math} can be modeled as linear functions with respect to the optimization variables, then the objective function becomes a quadratic function with integer variables.

\paragraph{Other Possible Objective Functions}
Apart from the objective functions mentioned above, the forms of QLMM can be optimized for various other objective functions as well. For instance, the purpose of optimization could be to find a QLMM with the lowest quantum circuit depth within the available qubit number limit. To achieve this, the qubit count in Eq.~\eqref{Low} can be set as a constraint to account for the qubit limit, while the quantum circuit depth represented by Eq.~\eqref{Fast} can be used as the objective function.

\subsubsection{Constraint of Optimized QLMM}
The constraints of the optimization problem ensure that a valid QLMM is constructed. Since the QLMM must be both consistent and stable, it must satisfy Eq.~\eqref{kconsistent} and the stability condition associated with Eq.~\eqref{sigma}. Additionally, to satisfy the conditions of the QLMM proposed in this paper, Eq.~\eqref{a0} must also be met. Depending on the implementation of summation, additional constraints related to the implementation must be satisfied. For example, if the summation is implemented using the method described in Appendix~\ref{ap1}, then Eq.~\eqref{A} must be satisfied. To prevent overflow in the floating-point representation, the inequality constraint
\begin{equation}\label{over}
\vb*{x}_{0}^{max}+\vb*{v}+(t_f-t_0) \vb*{u} < w_u(\vb*{E})
\end{equation}

\noindent must be satisfied, where $\vb*{x}_{0}^{max}$ represents the upper bounds of $\vb*{x}_0$ for every possible combination considered by the quantum algorithm, and $w_u(\vb*{E})$ is a function representing the upper bound of expressible floating-point numbers using $\vb*{E}$ qubits for the exponent representation. Similarly, to prevent underflow in the floating-point representation, the inequality constraint
\begin{equation}\label{under}
\vb*{x}_{0}^{min}+\vb*{v}+(t_f-t_0) \vb*{l} > w_l(\vb*{E})
\end{equation}
\noindent must be satisfied, where $\vb*{x}_{0}^{min}$ represents the lower bounds of $\vb*{x}_0$ for every possible combination considered by the quantum algorithm, and $w_l(\vb*{E})$ is a function representing the lower bound of expressible floating-point numbers using $\vb*{E}$ qubits for the exponent representation. Because the qubit count exhibits exponential trends, the optimization problem includes nonlinear constraints.

The error of the numerical solution during one time step includes local truncation error $\tau_h$ due to the limitation of the size of the time step, and round-off error due to the limitation of qubit number. The QLMM aims to bound the error generated through calculations in the quantum circuit to less than $\epsilon$. This condition is provided in
\begin{equation}\label{error}
(x_{0,d}^{max} +v_d+ (t_f-t_0) u_d)2^{-M_d} + \tau_h <\epsilon,  
\end{equation}
\noindent where $d=0,\cdots,D-1$.

Lastly, constraints specific to the oracle application can be added. For instance, the size of the time step can have an upper limit set to a particular value to ensure accurate oracle application. The constraints for the optimized QLMM include both linear and nonlinear functions with integer variables.

\subsubsection{Solving Optimized QLMM}

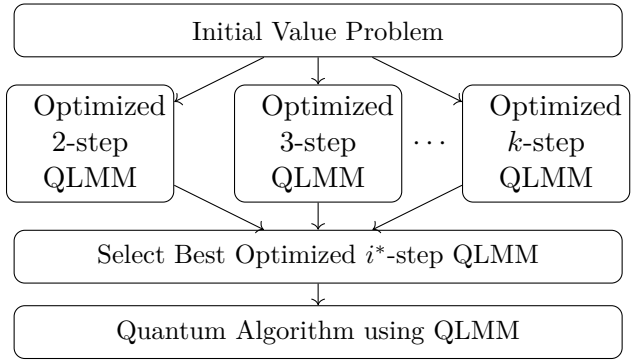
\begin{figure}[h]
\begin{center}
\begin{tikzpicture}
\node (start) [process] {Initial Value Problem};
\node (pro2) [process_s, below of =start, yshift = -0.5cm] {Optimized 3-step QLMM};
\node (pro1) [process_s, left of =pro2  , xshift = -2cm] {Optimized 2-step QLMM};
\node (pro3) [process_s, right of =pro2  , xshift = 2cm] {Optimized $k$-step QLMM};

\node at (1.5, -1.5) {$\cdots$};
\node (sel) [process, below of =pro2, yshift = -0.5cm] {Select Best Optimized $i^*$-step QLMM};
\node (oracle) [process, below of =sel] {Quantum Algorithm using QLMM};

\draw [->] (start) -- (pro1);
\draw [->] (start) -- (pro2);
\draw [->] (start) -- (pro3);
\draw [->] (pro1) -- (sel);
\draw [->] (pro2) -- (sel);
\draw [->] (pro3) -- (sel);
\draw [->] (sel)  -- (oracle);
\end{tikzpicture}
\end{center}
\caption{Sequence for efficiently using QLMM to solve a given IVP in quantum algorithms}\label{algorithm_fig}
\end{figure}

To efficiently use the QLMM for solving a given IVP under the objective functions defined in Eq.~\eqref{Low} or Eq.~\eqref{Fast}, the following process must be completed: 

First, compare the performance of the optimized QLMMs, ranging from 2-step to $k$-step, under constraints tailored for a specific purpose. As the optimization problem involves both integer and nonlinear constraints, it is classified as a non-convex MINLP. Solve this non-convex MINLP to find the optimal form of each $i$-step QLMM using the spatial branch and bound method \cite{9_2}. In a subroutine of this method, the outer approximation (OA) method \cite{9_3} can be used for convex MINLP problems. Employ optimization tools such as CPLEX \cite{9_5} or Gurobi Optimizer \cite{9_1} to address the problem. Each $i$-step QLMM yields different objective values within the same error range, and the optimal $i^*$-step QLMM is selected based on the lowest objective value. Finally, implement the quantum circuit using the best QLMM and integrate it into the quantum algorithm. The entire process is depicted in Fig.~\ref{algorithm_fig}.

\section{Simulation Result}
In this section, we verify that the QLMM can solve a simple example of an IVP using classical computer simulations.

\subsection {Computer Simulation using QLMM}

\paragraph {Purpose of Simulation}
Our simulation process is outlined as follows. First, a simple IVP with a system of differential equations is defined. Then, the optimized form of QLMM described in section~\ref{3_3} is found. Finally, the IVP is solved using QLMM, confirming both the solution of the IVP and the results obtained by applying an oracle to this solution.

\paragraph {Simulation Environment}
The non-convex MINLP problem for finding the optimized QLMM form was solved using Gurobi Optimizer 11.01 on a personal computer. The Gurobi Optimizer efficiently solves optimization problems through spatial branching \cite{9_1}. Additionally, QLMM computation results were verified by performing calculations using Python to simulate bit-level operations. Also, summation is assumed to be performed by the method from Appendix~\ref{ap1}.


\subsection {Oracle using Solution at Final Time}

\subsubsection {Problem Definition}
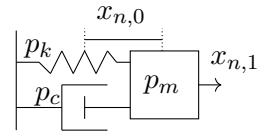
\begin{figure}[h]
\begin{center}
\begin{tikzpicture}[scale=0.60]
\draw (0,0) -- (0,2.2);
\draw (0,0.5) -- (1,0.5);
\draw (0.7, 0.75) node(x)  {\it{$p_c$}};
\draw (1,0) -- (1,1);
\draw (1,0) -- (2,0);
\draw (1,1) -- (2,1);
\draw (1.5,0.25) -- (1.5,0.75);
\draw (1.5,0.5) -- (2.5,0.5);
\draw (0,1.5) -- (0.5, 1.5);
\draw (0.5, 1.8) node(x)  {\it{$p_k$}};
\draw (0.5, 1.5) -- (0.75, 1.25);
\draw (0.75, 1.25) -- (1, 1.75);
\draw (1, 1.75) -- (1.25, 1.25) ;
\draw (1.25, 1.25) -- (1.5, 1.75);
\draw (1.5, 1.75) -- (1.75, 1.25);
\draw (1.75, 1.25) -- (2, 1.75);
\draw (2, 1.75) -- (2.2, 1.5);
\draw (2.2, 1.5) -- (2.5, 1.5);
\draw (2.5,0.25) rectangle (4,1.75);
\draw (3.2, 1) node(x)  {\it{$p_m$}};
\draw[densely dotted] (3.2,1.75) -- (3.2, 2.25);
\draw[densely dotted] (1.5,1.75) -- (1.5, 2.25);
\draw (1.5, 2) -- (3.2, 2);
\draw (2.3, 2.5) node(x)  {\it{$x_{n,0}$}};
\draw[->] (4,1) -- (4.5, 1);
\draw (4.8, 1.5) node(x)  {\it{$x_{n,1}$}};

\end{tikzpicture}
\end{center}
\caption{Spring-mass-damper system at $n$th time step}
\label{mck_fig}
\end{figure}

The spring-mass-damper system, depicted in Fig.~\ref{mck_fig}, consists of a mass connected to a fixed wall by a spring and a damper. 

\begin{figure*}[ht] 
    \begin{subfigure}{0.33\textwidth}
        \includegraphics[width=0.9\linewidth]{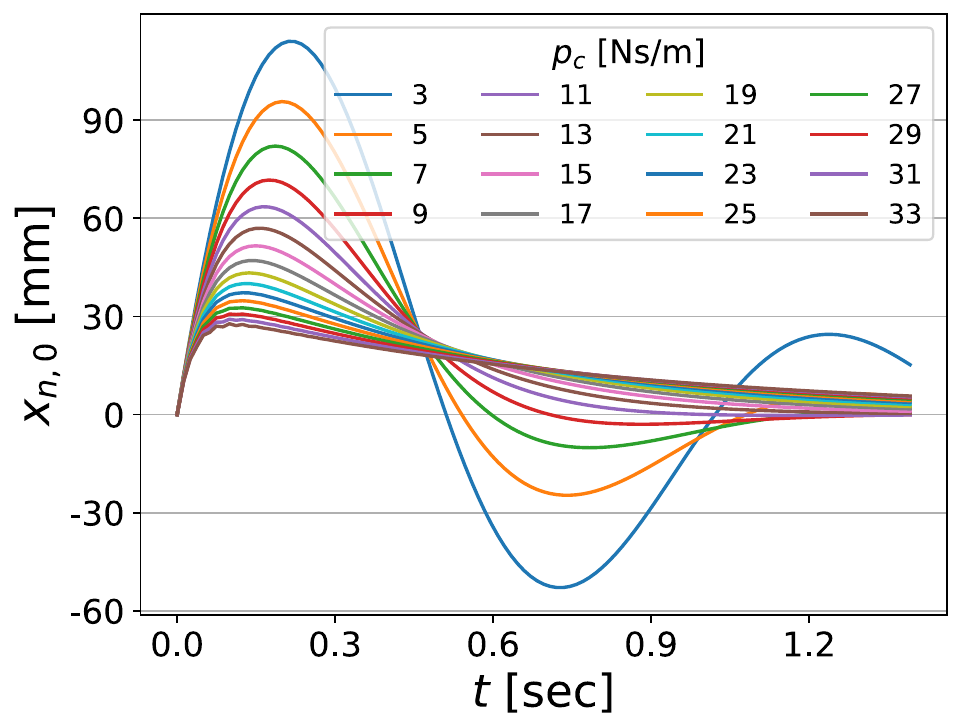}
        \caption{}
        \label{mck}
    \end{subfigure}%
    \begin{subfigure}{0.33\textwidth}
        \includegraphics[width=0.9\linewidth]{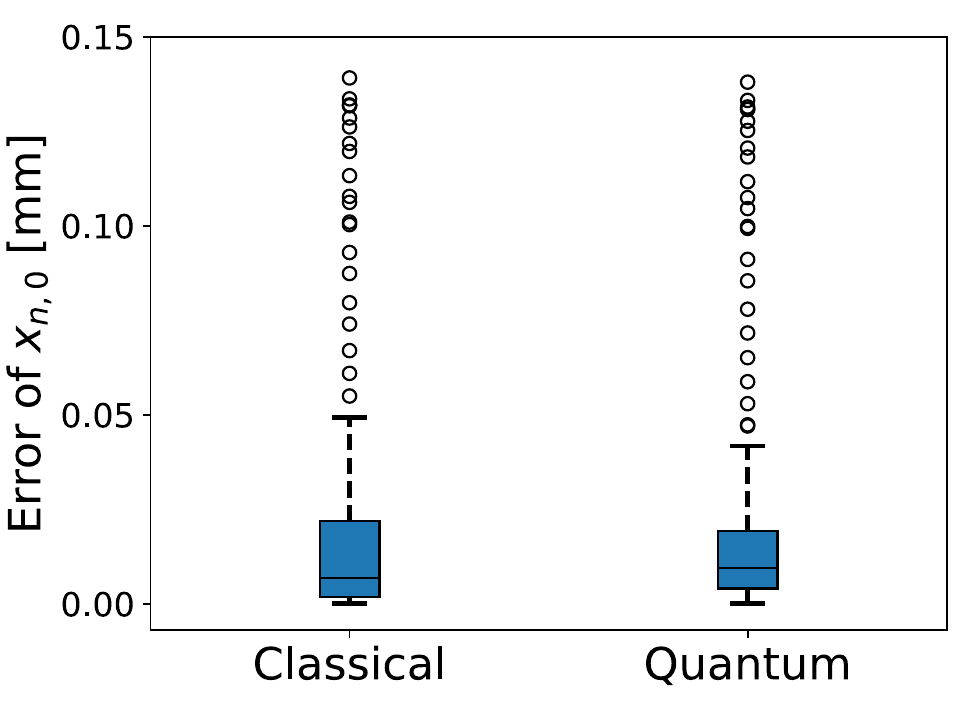}
        \caption{}
        \label{mck_error}
    \end{subfigure}
    \begin{subfigure}{0.33\textwidth}
        \includegraphics[width=0.9\linewidth]{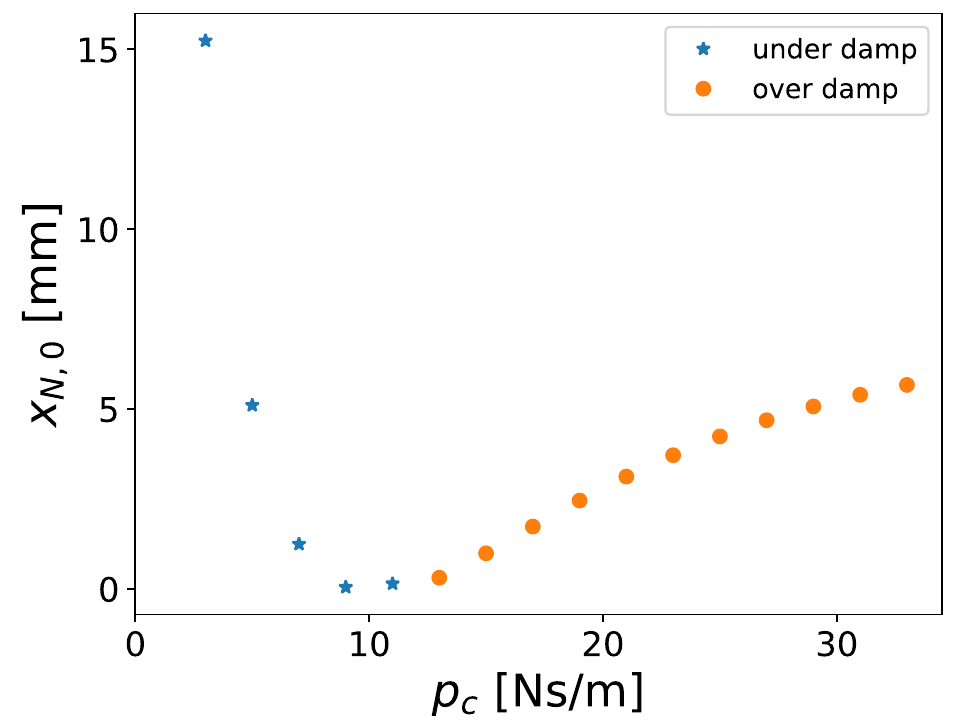}
        \caption{}
        \label{mck_oracle}
    \end{subfigure}%
    \caption{Simulation results for calculating the critical damping coefficient using QLMM (a) Numerical results of the location over time for different damping coefficients using QLMM. (b) Error in numerical results of the mass position over time using the 3-step LMM and 3-step QLMM for $p_c=13$. (c) Location of the mass at the final time for different damping coefficients.}
\end{figure*}

\noindent Here, $x_{n,0}$ and $x_{n,1}$ denote the location and velocity, respectively,  of the mass at the $n$th time step. Additionally, in this problem, the mass is represented by \begin{math}p_m\end{math}, the spring constant by \begin{math}p_k\end{math} and the damping coefficient \begin{math}p_c\end{math}. The IVP for this system is represented as
\begin{equation}\label{mck_model}\begin{split}
    &\vb*{x}_n\in \mathbb{R}^2, \vb*{x}_0 = [0, 1]^T,\\
&\vb*{f}_n=\begin{bmatrix}
   0 & 1 \\
   -p_k/p_m & -p_c/p_m
\end{bmatrix}\begin{bmatrix}
   x_{n,0} \\
   x_{n,1}
\end{bmatrix}_.
\end{split}\end{equation}

In this example, the task is to determine the critical damping by varying $p_c$ while keeping the $p_m$ and $p_k$ constant. For each $p_c$, QLMM computes the mass position at $t_f=1.4$ \si{s}. The damping coefficient that yields the smallest positive position, without the mass ever having a negative position, is identified as the critical damping coefficient. In the simulation, \begin{math} p_m=1\end{math} \si{kg}, \begin{math}p_k=40\end{math} \si{N.m^{-1}}, and \begin{math} p_c \end{math} varies from 3 to 33 \si{N.s.m^{-1}} across 16 cases. Theoretically, the damping coefficient for critical damping is \begin{math} 2\sqrt{p_m p_k} \end{math}, such that it equals 12.65\si{N.s.m^{-1}}. Both 2-step and 3-step QLMM are individually optimized, and the better one is selected for use.

\paragraph{Quantum Oracle Implementation}
To find the critical damping coefficient of the spring-mass-damper system, D\"{u}rr and H{\o}yer's algorithm can be applied, requiring the preparation of quantum oracle. The oracle must check the sign of each $x_{n,0}$ for $n=0$ to $N$ by comparing $y_{n,0}$ with $v_0$. Additionally, it compares $y_{N,0}$ with a threshold at each iteration step of D\"{u}rr and H{\o}yer's algorithm.

\subsubsection {Numerical Result from QLMM}

\paragraph {Form of Optimized QLMM}
To use a smaller number of qubits for QLMM implementation, the optimization results obtained with Gurobi were most efficient for a 3-step QLMM. Its form is given by
\begin{equation}\label{mck_result}\begin{split}
    &\vb*{y}_{n+3} +0.2427 \vb*{y}_{n+2} -0.7427 \vb*{y}_{n+1} -0.5 \vb*{y}_n\\
    &=  0.01243 ( 1.8714 \vb*{f}_{n+2} + 0.8714 \vb*{f}_{n+1}) 
\end{split}\end{equation}

\noindent and variables shown in Table~\ref{mck_table}. Because the number of $A_d$ qubits is consumed at every time step and the total number of time steps is large, the optimal solution is to minimize $A_d$, which is 1.

\begin{table}[h]
\begin{center}
\caption{Optimal constants of QLMM for finding the critical damping coefficient}
\label{mck_table}
\begin{tabular}{|c || c c c c|} 
 \hline
 $d$  & $E_d$ & $M_d$ & $A_d$ & $v_d$ \\ 
 \hline\hline
 0 & 3 & 25 & 1 & 10\\ 
 \hline
 1 & 4 & 27 & 1 & 69.7\\ 
 \hline
\end{tabular}
\end{center}
\end{table}

\paragraph {Numerical Result from QLMM}
The results from applying the 3-step QLMM, as in Eq.~\eqref{mck_result}, using Python are shown in Fig.~\ref{mck}. Fig.~\ref{mck_error} is a box-and-whisker diagram illustrating the errors of the 3-step LMM and the 3-step QLMM with respect to the analytic solutions. Since the QLMM calculation method differs only in bit count, it achieves the same level of accuracy as the 3-step LMM. The solution of the IVP obtained through QLMM closely resembles the analytical result.

\paragraph {Expected Result using Quantum Oracle}
Figure~\ref{mck_oracle} reconstructs the solutions expressed in Fig.~\ref{mck} at the final time step across different damping coefficients. By using these values in the quantum oracle, D\"{u}rr and H{\o}yer's algorithm will identify that $p_c=13 \si{N.s.m^{-1}}$, as the solution has the minimum value and is not an under-damped case, which agrees with analytical predictions. This confirms that QLMM can identify the damping coefficient that achieves critical damping for this system.

\begin{figure}[h]
  \centering
  \includegraphics[width=0.9\linewidth]{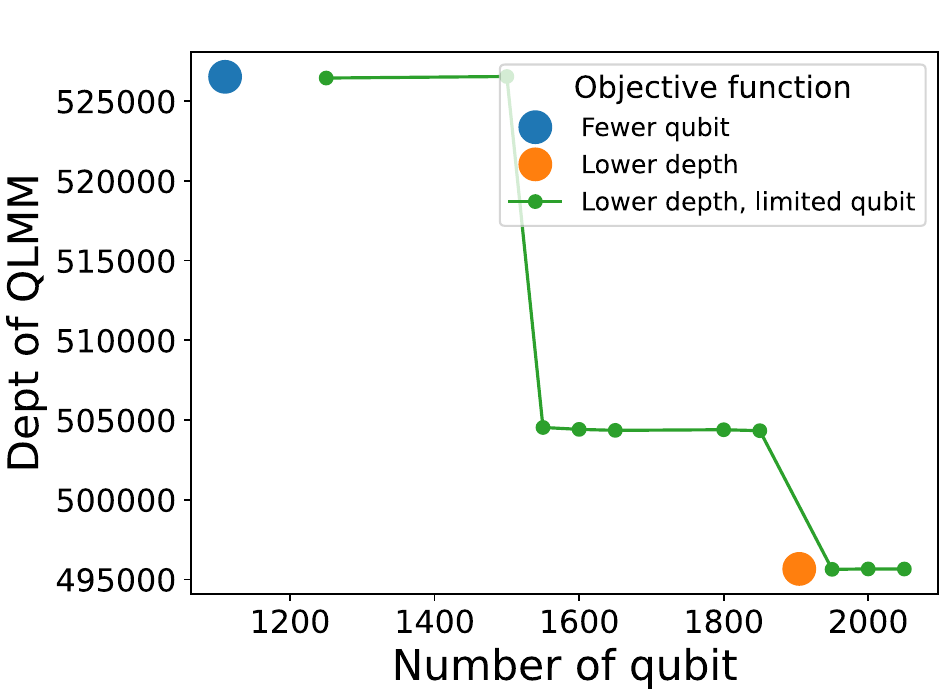}
  \caption{Qubits used vs depth of the circuit in different types of optimized QLMM}\label{mck_version}
\end{figure}

\paragraph {Comparing Different Types of Objective Functions}
Different types of objective functions can be applied to the spring-mass-damper system. Fig.~\ref{mck_version} shows the required number of qubits and the depth of the QLMM circuit from different perspectives. When circuit depth is minimized with a limit on the number of qubits, the solutions marked in green are obtained. Since the optimization problem is discrete, it converges to a specific number of qubits. The fewer qubit QLMM approach, which minimizes the number of qubits, is optimized to use the fewest possible qubits, whereas the lower depth QLMM approach is designed to achieve the smallest circuit depth.

\subsection {Oracle using Solutions at All Time Steps}

\subsubsection {Problem Definition}
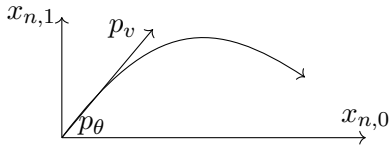
\begin{figure}[h]
\begin{center}
\begin{tikzpicture}[scale=0.80]
\draw[->] (0,0) -- (5,0);
\draw (5,0.3) node(x)  {$x_{n,0}$};

\draw[->] (0,0) -- (0,2);
\draw (-0.5,2) node(x)  {$x_{n,1}$};

\draw[->] (0,0) -- (1.5, 1.8);
\draw (1, 1.8) node(x)  {\it{$p_v$}};

\draw[->] (0,0) .. controls (1.5, 2) and (2.5,2) .. (4,1);
\draw (0.5, 0.2) node(x)  {$p_\theta$};

\end{tikzpicture}
\end{center}
\caption{Ballistic trajectory at $n$th time step}\label{canon_fig}
\end{figure}

We solve the problem of ballistic trajectories, as depicted in Fig.~\ref{canon_fig}.

\noindent Here, $x_{n,0}$ and $x_{n,1}$ denote the horizontal and vertical locations, respectively, and $x_{n,2}$ denotes the vertical velocity, at the $n$th time step. Additionally, in this problem, the object is launched at velocity $p_v$ with launch angle \begin{math} p_{\theta} \end{math}. The IVP for this system is represented as
\begin{equation}\label{canon_eq}\begin{split}
    &\vb*{x}_n\in \mathbb{R}^3, \vb*{x}_0=[0, 0, p_v \sin p_{\theta}]^T\\
    &\vb*{f}_n=[p_v \cos p_{\theta}, x_{n,2}, -9.8]_.
\end{split}\end{equation}

The task is to find the launch angle \begin{math} p_{\theta} \end{math} that results in the maximum distance when the projectile returns to its initial height. Theoretically, the angle \begin{math} p_{\theta} \end{math} that achieves the maximum distance is 45 degrees. In this simulation, \begin{math} p_v=40\end{math} \si{m.s^{-1}}, and the $p_{\theta}$ varies from 31 to 63 degrees in 16 cases. Both the 2-step and 3-step QLMM are optimized individually, and the best one is selected.

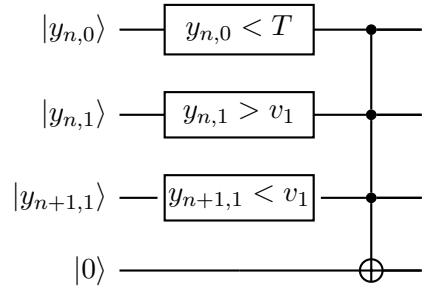
\begin{figure}[h]
\begin{center}
\begin{quantikz}[]
\lstick{$\ket{y_{n,0}}$} &\gate[style={inner xsep=18pt}]{y_{n,0} < T} &\ctrl{1} &\qw \qw\\
\lstick{$\ket{y_{n,1}}$} &\gate[style={inner xsep=18pt}]{y_{n,1}>v_1} &\ctrl{1} &\qw\qw\\
\lstick{$\ket{y_{n+1,1}}$}&\gate[style={inner xsep=18pt}]{y_{n+1,1}<v_1} &\ctrl{1} &\qw\qw\\
\lstick{$\ket{0}$} &\qw     &\targ{} &\qw\qw\\
\end{quantikz}
\end{center}
\caption{Implementation of the quantum oracle in D\"{u}rr and H{\o}yer's algorithm at the threshold value $T$ for this iteration} 
\label{Canon_Oracle}
\end{figure}

\paragraph{Quantum Oracle Implementation}
To determine the optimal launch angle for a ballistic trajectory, D\"{u}rr and H{\o}yer's algorithm can be applied, requiring the preparation of a quantum oracle. This oracle needs to find the time step $n$ when it returns to the ground by checking whether \begin{math} x_{n,1} \end{math} is positive and \begin{math} x_{n+1, 1} \end{math} is negative simultaneously. Additionally, it compares \begin{math} y_{n,0} \end{math} with the threshold value $T$ at each iteration step of D\"{u}rr and H{\o}yer's algorithm. Quantum oracle implementation for ballistic trajectories can be expressed as shown in Fig.~\ref{Canon_Oracle}.

\begin{figure*}[ht]
    \begin{subfigure}{0.33\textwidth}
        \includegraphics[width=0.9\linewidth]{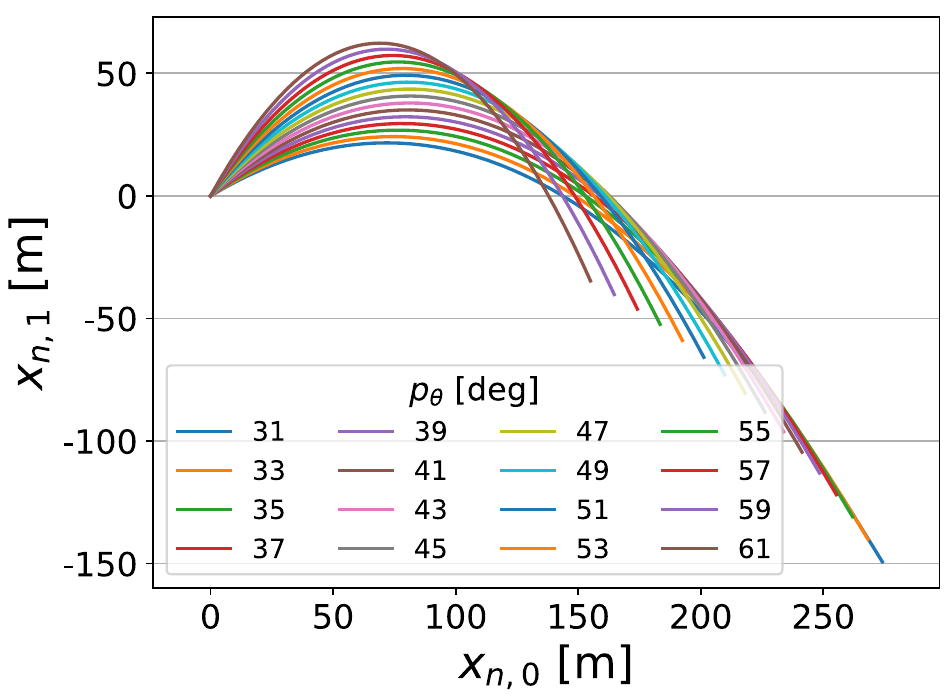}
        \caption{}
        \label{canon}
    \end{subfigure}%
    \begin{subfigure}{0.33\textwidth}
        \includegraphics[width=0.9\linewidth]{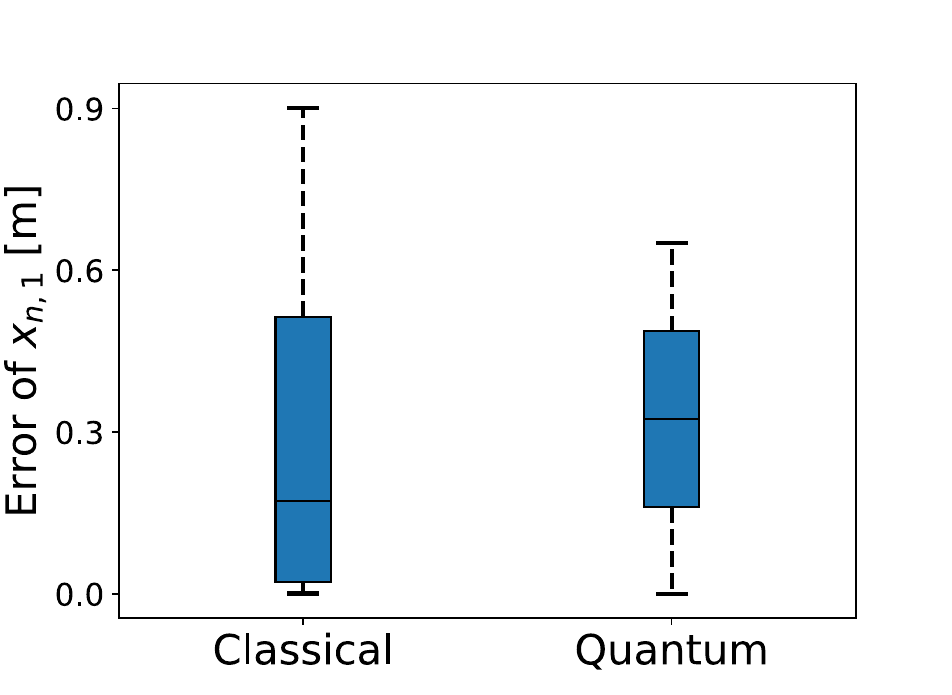}
        \caption{}
        \label{canon_error}
    \end{subfigure}
    \begin{subfigure}{0.33\textwidth}
        \includegraphics[width=0.9\linewidth]{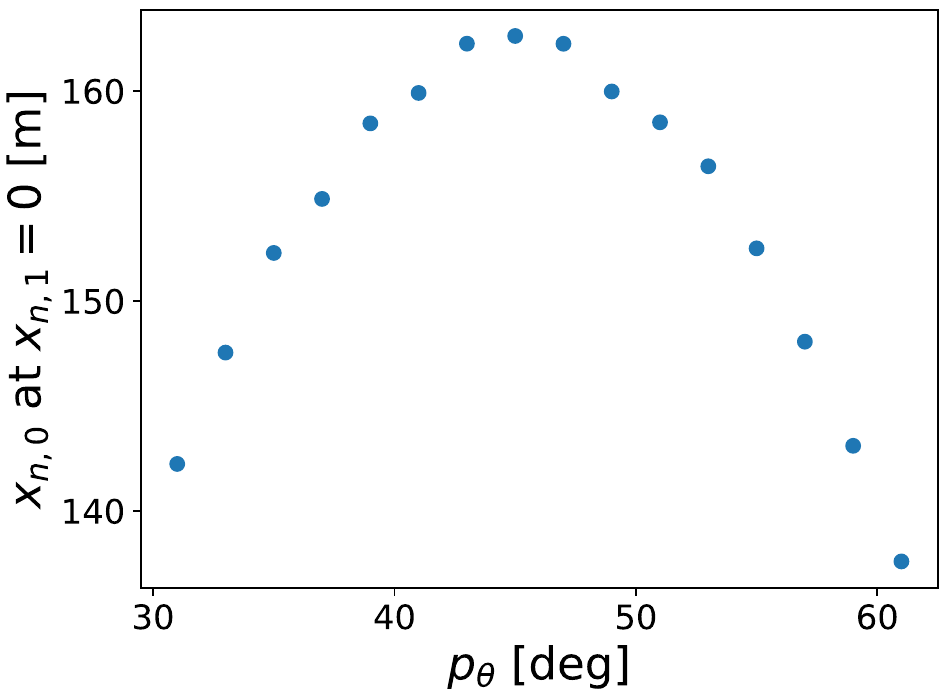}
        \caption{}
        \label{canon_oracle}
    \end{subfigure}%
    \caption{Simulation results for calculating ballistic trajectory using QLMM (a) Numerical results of the ballistic trajectories for different launch angles using QLMM. (b) Error in numerical results of the 2-dimensional trajectory for a launch angle($p_\theta$) of 45 degrees using the 2-step LMM and 2-step QLMM. (c) Horizontal location in the parabolic trajectory at height 0 for different launch angles.}
\end{figure*}

\subsubsection {Numerical Result from QLMM }

\paragraph {Form of Optimized QLMM}
To use a smaller number of qubits for implementing QLMM, the optimization results obtained with Gurobi were most efficient for a 2-step QLMM. When the time step $h$ is limited to 0.05 seconds to obtain a precise result from the oracle, the optimized form is given by
\begin{equation}\label{canon_result}
\vb*{y}_{n+2} - 1/2 \vb*{y}_{n+1} - 1/2 \vb*{y}_n = 0.05 \times 1.5 \vb*{f}_{n+1},
\end{equation}

\noindent and the relevant variables are shown in Table~\ref{canon_table}.

\begin{table}[h]
\begin{center}
\caption{Optimal constants of QLMM for finding the best launch angle of a ballistic trajectory}
\label{canon_table}
\begin{tabular}{|c || c c c c|} 
 \hline
 $d$  & $E_d$ & $M_d$ & $A_d$ & $v_d$ \\ 
 \hline\hline
 0 & 4 & 16 & 1 & 16.0\\ 
 \hline
 1 & 5 & 18 & 1 & 663.0\\ 
 \hline
 2 & 4 & 14 & 1 & 56.2\\ 
 \hline
\end{tabular}
\end{center}
\end{table}

\paragraph {Numerical Result from QLMM}
Fig.~\ref{canon} presents the ballistic trajectory for different launch angles using the 2-step QLMM. Fig.~\ref{canon_error} is a box-and-whisker diagram illustrating the errors of the 2-step LMM and the 2-step QLMM with respect to the analytic solutions. The results demonstrate that QLMM accurately reproduces the trajectory.

\paragraph {Expected Result using Quantum Oracle}
Fig.~\ref{canon_oracle} plots the horizontal range of a ballistic trajectory at different launch angles when the vertical height reaches zero. By using these values in the quantum oracle, D\"{u}rr and H{\o}yer's algorithm will identify that the ballistic trajectory at 45 degrees achieves the maximum range, which agrees with analytical predictions. This confirms that QLMM successfully identifies the angle yielding the maximum range for the ballistic trajectory.


\section{Conclusion}
This study introduced the quantum linear multistep method, as a method to efficiently solve initial value problems on quantum circuits for use in quantum oracle-based algorithms. Additionally, we presented a method to design QLMM to optimize qubit usage for specific problems. Through computer simulations, we demonstrated that the optimized QLMM can accurately solve IVPs, and the obtained solutions can be applied to problems such as the spring-mass-damper system and the ballistic trajectories.

The QLMM presented in this paper has broad applicability. It can be integrated into quantum oracles, used in search problems with Grover’s search algorithm, and applied to optimization problems using  D\"{u}rr and H{\o}yer's algorithm. If IVP problems can be efficiently encoded in quantum circuit, they could potentially achieve a quadratic speed-up in terms of combinations by using QLMM.


\section{Acknowledgment}
\label{sec:version}
This work was supported by Hyundai Motor Company and KIA Corporation, and also by the National Research Foundation of Korea (NRF) grant funded by the Korean government (MSIT) (No. RS-2024-00442855, No. RS-2024-00413957).

\bibliography{ref}

\begin{thebibliography}{10}

\bibitem{0_0}
G.~Wanner E.~Hairer, S. P.~Nørsett.
\newblock ``Solving ordinary differential equations i: Nonstiff problems''.
\newblock \href{https://dx.doi.org/https://doi.org/10.1007/978-3-540-78862-1}{Chapter~1}.
\newblock Springer. Berlin Heidelberg~(1993).
\newblock 2nd edition.

\bibitem{0_1}
M.~I. Davidzon.
\newblock ``Newton’s law of cooling and its interpretation''.
\newblock \href{https://dx.doi.org/10.1016/j.ijheatmasstransfer.2012.03.035}{International Journal of Heat and Mass Transfer {\bf 55}, 5397}~(2012).

\bibitem{0_2}
P.W. Shor.
\newblock ``Algorithms for quantum computation: discrete logarithms and factoring''.
\newblock In Proceedings 35th Annual Symposium on Foundations of Computer Science.
\newblock \href{https://dx.doi.org/10.1109/sfcs.1994.365700}{Page 124}.
\newblock SFCS-94. IEEE Comput. Soc. Press~(1994).

\bibitem{0_3}
A.~W. Harrow, A.~Hassidim, and S.~Lloyd.
\newblock ``Quantum algorithm for linear systems of equations''.
\newblock \href{https://dx.doi.org/10.1103/physrevlett.103.150502}{Physical Review Letters {\bf 103}, 150502}~(2009).

\bibitem{1_2}
A.~M. Childs, R.~Kothari, and R.~D. Somma.
\newblock ``Quantum algorithm for systems of linear equations with exponentially improved dependence on precision''.
\newblock \href{https://dx.doi.org/10.1137/16m1087072}{SIAM Journal on Computing {\bf 46}, 1920}~(2017).

\bibitem{2}
L.~K. Grover.
\newblock ``A fast quantum mechanical algorithm for database search''.
\newblock In Proceedings of the twenty-eighth annual ACM symposium on Theory of computing - STOC ’96.
\newblock \href{https://dx.doi.org/10.1145/237814.237866}{Page 212}.
\newblock STOC ’96. ACM Press~(1996).

\bibitem{3}
C.~D\"{u}rr and P.~H{\o}yer.
\newblock ``A quantum algorithm for finding the minimum''~(1999).
\newblock  \href{http://arxiv.org/abs/quant-ph/9607014}{arXiv:quant-ph/9607014}.

\bibitem{4_1}
D.~W. Berry.
\newblock ``High-order quantum algorithm for solving linear differential equations''.
\newblock \href{https://dx.doi.org/10.1088/1751-8113/47/10/105301}{Journal of Physics A: Mathematical and Theoretical {\bf 47}, 105301}~(2014).

\bibitem{1}
D.~W. Berry, A.~M. Childs, A.~Ostrander, and G.~Wang.
\newblock ``Quantum algorithm for linear differential equations with exponentially improved dependence on precision''.
\newblock \href{https://dx.doi.org/10.1007/s00220-017-3002-y}{Communications in Mathematical Physics {\bf 356}, 1057}~(2017).

\bibitem{4}
B.~Zanger, C.~B. Mendl, M.~Schulz, and M.~Schreiber.
\newblock ``Quantum algorithms for solving ordinary differential equations via classical integration methods''.
\newblock \href{https://dx.doi.org/10.22331/q-2021-07-13-502}{Quantum {\bf 5}, 502}~(2021).

\bibitem{2_0}
L.~K. Grover.
\newblock ``Fixed-point quantum search''.
\newblock \href{https://dx.doi.org/10.1103/physrevlett.95.150501}{Physical Review Letters {\bf 95}, 150501}~(2005).

\bibitem{2_0_1}
T.~J. Yoder, G.~H. Low, and I.~L. Chuang.
\newblock ``Fixed-point quantum search with an optimal number of queries''.
\newblock \href{https://dx.doi.org/10.1103/physrevlett.113.210501}{Physical Review Letters {\bf 113}, 210501}~(2014).

\bibitem{2_1}
A.~Ambainis.
\newblock ``Polynomial degree and lower bounds in quantum complexity: Collision and element distinctness with small range''.
\newblock \href{https://dx.doi.org/10.4086/toc.2005.v001a003}{Theory of Computing {\bf 1}, 37}~(2005).

\bibitem{2_2}
M.~Udrescu, L.~Prodan, and M.~Vlăduţiu.
\newblock ``Implementing quantum genetic algorithms: a solution based on grover’s algorithm''.
\newblock In Proceedings of the 3rd conference on Computing frontiers.
\newblock \href{https://dx.doi.org/10.1145/1128022.1128034}{Page~71}.
\newblock CF06. ACM~(2006).

\bibitem{2_3}
L.~Hsu.
\newblock ``Quantum secret-sharing protocol based on grover’s algorithm''.
\newblock \href{https://dx.doi.org/10.1103/physreva.68.022306}{Physical Review A {\bf 68}, 022306}~(2003).

\bibitem{3_1}
A.~Shukla and P.~Vedula.
\newblock ``Trajectory optimization using quantum computing''.
\newblock \href{https://dx.doi.org/10.1007/s10898-019-00754-5}{Journal of Global Optimization {\bf 75}, 199}~(2019).

\bibitem{3_2}
A.~Malossini, E.~Blanzieri, and T.~Calarco.
\newblock ``Quantum genetic optimization''.
\newblock \href{https://dx.doi.org/10.1109/TEVC.2007.905006}{IEEE transactions on evolutionary computation {\bf 12}, 231}~(2008).

\bibitem{5_2}
W.~E. Boyce, R.~C. DiPrima, and D.~B. Meade.
\newblock ``Elementary differential equations and boundary value problems''.
\newblock Chapter 1,8.
\newblock John Wiley \& Sons. ~(2021).
\newblock 9th edition.

\bibitem{5_0}
D.~F. Griffiths and D.~J. Higham.
\newblock ``Numerical methods for ordinary differential equations: initial value problems''.
\newblock \href{https://dx.doi.org/10.1007/978-0-85729-148-6}{Chapter 4-7}.
\newblock Springer. London~(2010).

\bibitem{5_1_1}
G.~Dahlquist.
\newblock ``Convergence and stability in the numerical integration of ordinary differential equations''.
\newblock \href{https://dx.doi.org/10.7146/math.scand.a-10454}{MATHEMATICA SCANDINAVICA {\bf 4}, 33}~(1956).

\bibitem{5_1_2}
G.~G. Dahlquist.
\newblock ``A special stability problem for linear multistep methods''.
\newblock \href{https://dx.doi.org/10.1007/bf01963532}{BIT {\bf 3}, 27}~(1963).

\bibitem{5_3}
K.E. Atkinson.
\newblock ``An introduction to numerical analysis''.
\newblock Chapter~6.
\newblock John Wiley \& Sons. ~(1978).
\newblock 2nd edition.

\bibitem{7_3}
L.~Ruiz-Perez and J.~C. Garcia-Escartin.
\newblock ``Quantum arithmetic with the quantum fourier transform''.
\newblock \href{https://dx.doi.org/10.1007/s11128-017-1603-1}{Quantum Information Processing {\bf 16}, 152}~(2017).

\bibitem{7_0}
T.~Haener, M.~Soeken, M.~Roetteler, and K.~M. Svore.
\newblock ``Quantum circuits for floating-point arithmetic''.
\newblock In Jarkko Kari and Irek Ulidowski, editors, Reversible Computation.
\newblock \href{https://dx.doi.org/https://doi.org/10.1007/978-3-319-99498-7_11}{Page 162}.
\newblock Springer International Publishing~(2018).

\bibitem{7_2}
S.~S. Gayathri, R.~Kumar, and S.~Dhanalakshmi.
\newblock ``Efficient floating-point division quantum circuit using newton-raphson division''.
\newblock \href{https://dx.doi.org/10.1088/1742-6596/2335/1/012058}{Journal of Physics: Conference Series {\bf 2335}, 012058}~(2022).

\bibitem{9_10}
S.~Burer and A.~N. Letchford.
\newblock ``Non-convex mixed-integer nonlinear programming: A survey''.
\newblock \href{https://dx.doi.org/10.1016/j.sorms.2012.08.001}{Surveys in Operations Research and Management Science {\bf 17}, 97}~(2012).

\bibitem{9_6}
S.~Boyd and L.~Vandenberghe.
\newblock ``Convex optimization''.
\newblock \href{https://dx.doi.org/10.1017/cbo9780511804441}{Chapter III}.
\newblock Cambridge University Press. ~(2004).

\bibitem{9_7}
P.~Bonami, L.~T. Biegler, A.~R. Conn, G.~Cornuéjols, I.~E. Grossmann, C.~D. Laird, J.~Lee, A.~Lodi, F.~Margot, N.~Sawaya, and A.~Wächter.
\newblock ``An algorithmic framework for convex mixed integer nonlinear programs''.
\newblock \href{https://dx.doi.org/10.1016/j.disopt.2006.10.011}{Discrete Optimization {\bf 5}, 186}~(2008).

\bibitem{9}
C.~A. Floudas.
\newblock ``Deterministic global optimization: theory, methods and applications''.
\newblock \href{https://dx.doi.org/https://doi.org/10.1007/978-1-4757-4949-6}{Volume~37, page 576}.
\newblock Springer Science \& Business Media. ~(2013).

\bibitem{9_2}
E.M.B. Smith and C.C. Pantelides.
\newblock ``A symbolic reformulation/spatial branch-and-bound algorithm for the global optimisation of nonconvex minlps''.
\newblock \href{https://dx.doi.org/10.1016/s0098-1354(98)00286-5}{Computers \& Chemical Engineering {\bf 23}, 457}~(1999).

\bibitem{9_3}
R.~Fletcher and S.~Leyffer.
\newblock ``Solving mixed integer nonlinear programs by outer approximation''.
\newblock \href{https://dx.doi.org/10.1007/bf01581153}{Mathematical Programming {\bf 66}, 327}~(1994).

\bibitem{9_5}
S.~Nickel, C.~Steinhardt, H.~Schlenker, and W.~Burkart.
\newblock ``Decision optimization with ibm ilog cplex optimization studio''.
\newblock \href{https://dx.doi.org/https://doi.org/10.1007/978-3-662-65481-1}{Springer}. Berlin, Heidelberg~(2022).

\bibitem{9_1}
Gurobi Optimization.
\newblock ``Gurobi optimizer reference manual''.
\newblock Gurobi Optimization.
\newblock 11.0 edition~(2024).

\end{thebibliography}
\bibliographystyle{quantum}


\onecolumn\newpage
\appendix

\section{Summation using Reduced Number of Auxiliary Qubit}\label{ap1}
\subsection {Quantum Circuit Implementation}

We introduce a quantum circuit implementation for adding two positive floating-point numbers using auxiliary qubits. The floating-point numbers $r_a$ and $r_b$ consist of mantissas $m_a$ and $m_b$ and exponents $e_a$ and $e_b$. Each number is represented using $E_a$ and $E_b$ qubits for the exponents and $M_a$ and $M_b$ qubits for the mantissas. The entire process, illustrated in Fig.~\ref{Add_fig}a, proceeds as follows:  

Initially, the exponent of $r_a$ is stored in a concatenated quantum register composed of $Q_1$ and $Q_0$, and the mantissa of $r_a$ is stored in $Q_6$, with the most significant bit (MSB) located in the lower part of the circuit. Similarly, the exponent and mantissa of $r_b$ are stored in quantum registers $Q_2$ and $Q_9$, respectively. The remaining quantum registers are initialized to the zero state. First, the difference between the exponents of $r_a$ and $r_b$ is calculated using the $g_-$ block whose function is depicted in Fig.~\ref{Add_fig}d and implementation is described in Ref.~\cite{7_3}. 
After subtraction, the combined quantum register ($Q_1$, $Q_0$) stores $e_a-e_b$ in two's complement form, with $Q_1$ serving as MSB.
 If $e_a < e_b$, $e_b-e_a$ is stored in $Q_3$ using the quantum circuit described in Fig.~\ref{Add_fig}b. If $e_b$ is larger, the mantissa $m_a$ in quantum register $Q_6$ is shifted into another combined quantum register ($Q_6$ and $Q_5$) using the $g_s$ circuit described in Ref.~\cite{7_0}. Note that the number of qubits inside $Q_5$ ($A$) should be bigger than $\log_2 (r_b/r_a)$.
Conversely, if $e_a$ is larger, the mantissa $m_b$ in $Q_9$ is shifted into the combined register ($Q_9$ and $Q_8$) using the same $g_s$ circuit to align the mantissas. Next, the adjusted mantissas of $r_a$ and $r_b$, stored in $Q_6$ and $Q_9$, are added using the $g_+$ circuit, with the result stored in the combined register ($Q_7$ and $Q_6$). After the addition, quantum register $Q_8$ is uncomputed. The $g_+$ gate is then applied to restore the exponent of $r_a$ by adding $Q_2$ to the combined registers ($Q_1$ and $Q_0$).
If $e_a < e_b$, the combined register ($Q_1$ and $Q_0$) is updated to $e_b$ by adding the value in $Q_3$ with the following $g_+$. If an overflow occurs during the mantissa calculation, $Q_7$ will be set to 1. To ensure that the summation result remains in the same registers as $r_a$, the overflow is handled by copying $Q_7$ into $Q_{10}$ and incrementing the exponents stored in the combined register ($Q_1$ and $Q_0$) by 1. Based on $Q_{10}$, the mantissa stored in the combined register ($Q_7$, $Q_6$, $Q_4$) are shifted. Finally, the summation result is stored in quantum registers $Q_0$, $Q_1$, and $Q_6$.

While some quantum registers return to the zero state, others retain arbitrary values. A total of $A + \lfloor\log_2{A}\rfloor + 3$ qubits cannot be uncomputed during this process. However, the values in $\ket{etc}$ can be uncomputed using the circuits illustrated in Fig.~\ref{final_time} or Fig.~\ref{every_time}.


\subsection {Number of Auxiliary Qubit Needed}
To perform the QLMM operation, it is necessary to add the terms $-\alpha_0 \vb*{y_n}$ and $\sum_{i=1}^{k-1} (-\alpha_i \vb*{y}_{n+i} + h \beta_i \vb*{f}_{n+i})$. Let us assume the summation is performed by the quantum circuit shown in Fig.~\ref{Add_fig}a, with $e_a = -\alpha_0 y_{n,d}$ and $e_b = \sum_{i=1}^{k-1} (-\alpha_i y_{n+i, d} + h \beta_i f_{n+i, d})$. In this case, the term $-\alpha_0 y_{n,d}$ has an $A$-qubit margin for expression, and the value that can be expressed considering this margin is $-2^A \alpha_0 y_{n,d}$. The term $\sum_{i=1}^{k-1} (-\alpha_i \vb*{y}_{n+i} + h \beta_i \vb*{f}_{n+i})$ has the same value as $\vb*{y}_{n+k} + \alpha_0 \vb*{y}_n$. The inequality can be written as $-(2^A+1)\alpha_0 \vb*{y_n} > \vb*{y}_{n+k}$. By using the upper bound of the derivative, the inequality can be expressed as
\begin{equation}\label{A}
(-\alpha_0(2^A+1)-1) \vb*{y_n} > hk\vb*{u},  \quad n=0,\cdots,N.
\end{equation}

To minimize the required auxiliary qubits, $A$ must be minimized. In the case where $A=0$, Eq.~\eqref{A} becomes $-(\alpha_0 + 1) \vb*{y}_n > hk \vb*{u}$ for all $n$. However, for QLMM, the left-hand term is negative, while the right-hand term is positive, making implementation impossible. Therefore, in general, $A$ must be greater than 0 and requires at least 1 qubit.

\begin{figure}[h]
\centering
\begin{tikzpicture}
\node at (0,-5.5) {(a)};

\node at (0, 0) {
\begin{quantikz}[column sep=0.1cm, row sep=0.2cm,transparent]
\lstick{$\ket{e_{a[2:E_a]}}_{Q_0}$} &\qwbundle{E_a-1}&\qw&\qw&\qw&\qw&\qw&\qw&\qw&\qw&\qw&\qw&\qw&\gate[3]{g_-}\gateinput[2]{0} &\gate[4]{(b)}&\qw &\gate[10]{g_{s2}}\gateinput{0}&\qw &\gate[10]{g_{s2}^\dag}\gateoutput{0} &\gate[3]{g_+}\gateinput[2]{0}&\gate[4]{g_+}\gateinput[2]{0}&\gate[2]{g_{+1}}  &\qw  &\qw & \rstick{$\ket{e_{a+b[2:E_a]}}$}\qw\\
\lstick{$\ket{e_{a[1]}}_{Q_1}$}     &\qw   &\qw&\qw&\qw&\qw&\qw&\qw&\qw&\qw&\qw&\qw&\qw&      &      &\ctrl{2}   &\gateinput{1}  &\qw            &\gateoutput{1}& &    &               &\qw&\qw  & \rstick{$\ket{e_{a+b[1]}}$}\qw\\
\lstick{$\ket{e_b}_{Q_2}$}          &\qwbundle{E_b}&\qw&\qw&\qw&\qw&\qw&\qw&\qw&\qw&\qw&\qw &\qw&\gateinput{1} &\linethrough&\qw        &\linethrough   &\qw            &\linethrough &\gateinput{1}&\linethrough&\qw            &\qw&\qw  & \rstick{$\ket{e_b}$}\qw\\
\lstick{$\ket{0}_{Q_3}$}            &\qwbundle{\lfloor\log_2{A}\rfloor +1}&\qw&\qw&\qw&\qw&\qw&\qw&\qw&\qw&\qw&\qw&\qw&\qw&      &\gate[4]{g_s}\gateinput{$\ket{s}$}&\linethrough &\qw            &\linethrough &\qw      &\gateinput{1}&\qw            &\qw&\qw  & \rstick{$\ket{etc}$}\qw\\
\lstick{$\ket{0}_{Q_4}$}            &\qw   &\qw&\qw&\qw&\qw&\qw&\qw&\qw&\qw&\qw&\qw&\qw  &\qw &\qw   &\linethrough&\linethrough  &\qw            &\linethrough &\qw      &\qw        &\qw            &\gate[7]{g_s}\gateinput[4]{$\ket{x}$}&\qw  & \rstick{$\ket{etc}$}\qw\\
\lstick{$\ket{0}_{Q_5}$}            &\qwbundle{A}   &\qw&\qw&\qw&\qw&\qw&\qw&\qw&\qw&\qw&\qw&\qw  &\qw &\qw   &\gateinput[2]{$\ket{x}$}&\linethrough   &\qw            &\linethrough &\qw      &\qw        &\qw            &\linethrough&\qw & \rstick{$\ket{etc}$}\qw\\
\lstick{$\ket{m_a}_{Q_6}$}          &\qwbundle{M_a} &\qw&\qw&\qw&\qw&\qw&\qw&\qw&\qw&\qw&\qw&\qw  &\qw &\qw   &           &\linethrough   &\gate[4]{g_+}\gateinput[2]{0}&\linethrough &\qw      &\qw        &\qw            &       &\qw & \rstick{$\ket{m_{a+b}}$}\qw\\
\lstick{$\ket{0}_{Q_7}$}            &\qw   &\qw&\qw&\qw&\qw&\qw&\qw&\qw&\qw&\qw&\qw&\qw  &\qw &\qw   &\qw        &\linethrough   &               &\linethrough &\ctrl{3} &\qw&\ctrl{-6}&       &\qw & \rstick{$\ket{0}$}\qw\\
\lstick{$\ket{0}_{Q_8}$}            &\qwbundle{M_b} &\qw&\qw&\qw&\qw&\qw&\qw&\qw&\qw&\qw&\qw&\qw  &\qw &\qw   &\qw        &\gateinput{2}  &\linethrough   &\gateoutput{2}&\qw      &\qw        &\qw            &\linethrough &\qw & \rstick{$\ket{0}$}\qw\\
\lstick{$\ket{m_b}_{Q_9}$}          &\qwbundle{M_b} &\qw&\qw&\qw&\qw&\qw&\qw&\qw&\qw&\qw&\qw&\qw  &\qw &\qw   &\qw        &\gateinput{3}  &\gateinput{1}  &\gateoutput{3}&\qw      &\qw        &\qw            &\linethrough &\qw & \rstick{$\ket{m_b}$}\qw\\
\lstick{$\ket{0}_{Q_{10}}$}         &\qw   &\qw&\qw&\qw&\qw&\qw&\qw&\qw&\qw&\qw&\qw&\qw  &\qw &\qw   &\qw        &\qw            &\qw            &\qw          &\targ{}  &\qw        &\qw            &\gateinput{$\ket{s}$} &\qw & \rstick{$\ket{etc}$}\qw\\
\end{quantikz}};
\node at (0, -10) {(b)};

\node at (1, -8) {
    \begin{quantikz}[column sep=0.1cm, row sep=0.5cm]
    \lstick{$\ket{e_{a[\gamma:E_a]}}$}&\qwbundle{\lfloor\log_2{A}\rfloor +1}&\qw&\qw&\qw&\qw&\qw&\qw&\qw&\qw&\qw&\qw    &\ctrl{2}   &\qw        &\qw     &\qw\\
    \lstick{$\ket{e_{a[2:\gamma-1]}}$}&\qwbundle{\gamma-2}                  &\qw&\qw&\qw&\qw&\qw&\qw&\qw&\qw&\qw&\qw    &\qw        &\qw        &\qw     &\qw\\
    \lstick{$\ket{e_{a[1]}}$}         &\qw                                  &\qw&\qw&\qw&\qw&\qw&\qw&\qw&\qw&\qw&\qw    &\ctrl{1}   &\ctrl{1}   &\ctrl{1}&\qw\\
    \lstick{$\ket{0}_{Q_3}$}          &\qw                                  &\qw&\qw&\qw&\qw&\qw&\qw&\qw&\qw&\qw&\qw    &\targ{}    &\targ{}    &\gate[1]{g_{+1}} &\rstick{ $\begin{cases}
    \ket{0}_{Q_3}, &\text{if } e_{a[1]}=0 \\
    \ket{e_b - e_a}_{Q_3}, &\text{if } e_{a[1]}=1
    \end{cases}$ }\qw\\
    \end{quantikz}};

\node at (-3, -14.5) {(c)};

\node at (-5, -13) {\begin{quantikz}
    &\gate[4]{g_{s2}}\gateinput{0} &\qw\\
    &\gateinput{1}  &\qw\\
    &\gateinput{2}  &\qw\\
    &\gateinput{3}  &\qw\\
    \end{quantikz}};
\node at (-3, -13) {=};
\node at (-1, -13) {\begin{quantikz}
    \lstick{1}&\octrl{1}                     &\qw\\
    \lstick{0}&\gate[3]{g_{s}}\gateinput{$\ket{s}$}  &\qw\\
    \lstick{2}&\gateinput[2]{$\ket{x}$}      &\qw\\
    \lstick{3}&                     &\qw\\
    \end{quantikz}
    };

\node at (4, -14.5) {(d)};

\node at (4, -13) {
    \begin{quantikz}
    \lstick{$\ket{n_a}$}    &\gate[2]{g_{\pm}}\gateinput{0}  &\rstick{$\ket{n_a \pm n_b}$}\qw\\
    \lstick{$\ket{n_b}$}    &\gateinput{1}                   &\rstick{$\ket{n_b}$}    \qw\\
    \end{quantikz}
    };

\end{tikzpicture}

\caption{(a) Quantum circuit for implementing floating-point summations using fewer auxiliary qubits. $g_{+1}$: Quantum circuit for incrementing by 1. $g_s$: Quantum circuit for shifting values to compensate for exponent differences, as described in Ref.~\cite{7_0}. (b) Quantum circuit for copying the difference of exponents when $e_a < e_b$, $\gamma=E_a-\lfloor\log_2{A}\rfloor$ (c) Alternative implementation of the quantum circuit $g_s$ to compensate the mantissa due to exponent differences. (d) Quantum circuit $g_\pm$ for the summation or subtraction of integers $n_a$ and $n_b$, as described in Ref.~\cite{7_3}.}

\label{Add_fig}
\end{figure}

\end{document}